%% file: nullamr_jcp.tex
\documentclass[amssymb]{elsart}
\usepackage{setspace}
\usepackage{graphicx}
\usepackage{psfrag}

\newcommand{\Screll}{{\mathcal L}}

\begin{document}
\begin{frontmatter}

\title{Adaptive Mesh Refinement for Characteristic Codes}

\author{Frans Pretorius$^{1}$ and Luis Lehner$^{2}$}

\address{$^{1}$ Theoretical Astrophysics 130-033\\
         California Institute of Technology,\\
         Pasadena, CA, 91125 \\
         $^{2}$ Department of Physics and Astronomy\\
         Louisiana State University, \\
         Baton Rouge, LA 70810.}

\begin{abstract}
The use of adaptive mesh refinement (AMR) 
techniques is crucial for accurate and efficient simulation of 
higher dimensional spacetimes. In this work
we develop an adaptive algorithm tailored to the integration of 
finite difference discretizations of wave-like
equations using characteristic coordinates. We demonstrate the
algorithm by constructing a code implementing the Einstein-Klein-Gordon
system of equations in spherical symmetry.
We discuss how the algorithm can trivially be 
generalized to higher dimensional systems, and suggest a method that 
can be used to parallelize a characteristic code.
\end{abstract}
\end{frontmatter}

\maketitle

\section{Introduction}
The investigation of the characteristic structure of a system of partial differential
equations (PDEs) gives valuable insight on the behavior of allowed solutions. Analytical
studies have long benefited from understanding and employing this knowledge. In the numerical
realm, efficient algorithms have been developed which exploit the underlying characteristic
structure. These algorithms are obtained by formally transforming to characteristic variables,
realizing how these need to be updated, and then employing this information to construct an algorithm
using the original variables. However, the explicit integration
along characteristics, where coordinates are chosen adapted to the characteristic directions,
has received considerably less attention. Indeed, this option has so far only been actively
pursued within general relativity (GR), although as discussed in\cite{winicour,lehner} there are
powerful reasons to consider this option in systems described by wave-like equations.

The characteristic formulation of general relativity has since its 
introduction in the 60's played an important
role as a tool to investigate different aspects of the theory 
(see for instance \cite{bondi,sachs,NP,tamb-win,jeffccm,todreviewtubiengen}.) 
A clean picture of the effect
of gravitational waves and their geometric manifestation, links between the structure of
future null infinity and some interior sources, and the analysis of singularity structure 
are some areas that have effectively been tackled by this approach. 

In recent years, the formalism has also displayed its usefulness in the numerical arena.
Investigations of critical phenomena\cite{hamadestewart,garfinkle,husain,burko,brady},
wave propagation on non-trivial backgrounds\cite{isaacson} and
simulations of spacetimes containing black holes\cite{hpgn,mbh,dinverno,bartnik} or neutron
stars\cite{papasiebel} have benefited from exploiting this approach.
Numerical simulations of black hole spacetimes are of considerable
importance for the detection and analysis of gravitational waves that could be measured
by the new generation of gravitational wave detectors. Among the systems expected to produce
gravitational waves of sufficient intensity are those containing black holes and neutron stars,
and it would be ideal if the success obtained simulating single black holes with 3D characteristic codes 
could be translated to these systems. Preliminary indications that this is likely the case for a subset 
of possible black hole-neutron star
binaries can be found in \cite{mattchar,papasiebel}, where higher dimensional, non-vacuum scenarios
(yet simpler than the 3D binary black hole-neutron star system) have been accurately simulated.

Unfortunately, the computational requirements for modeling a binary black hole-neutron star 
system are quite high if one of the goals 
is to predict the waveforms produced by it in a quantitative manner. Since the expected energy
output in gravitational waves is at most a few percent of the total mass of the 
system, the numerical simulation must
guarantee that any systematic (numerical) error is well below this target.
The major issue in achieving an accurate description of the system, when a stable discretization has been
obtained, is to adequately resolve the different length scales involved. These scales are naturally defined
by the stellar dynamics, the black hole, and the distant weak-field regime where the gravitational waves
are extracted. Covering all of these scales with a uniform grid (in a finite-difference based numerical
simulation) of sufficient resolution to accurately model the smallest features
is not only a waste of resources, but may be impossible to achieve on
contemporary computers. Therefore, we need to use techniques that better exploit available resources.
One such technique is {\it adaptive mesh refinement} (AMR), whereby the simulation can 
dynamically adjust the grid resolution in different regions of the domain to 
adequately resolve all features with sufficient, but not excess, resolution.

In this paper we investigate the use of AMR techniques for characteristic
evolution. Although we concentrate on the GR case, the algorithms presented here
can readily be applied to many other equations. The use of characteristic AMR in GR
has been partially addressed by several authors in the 
past\cite{hamadestewart,garfinkle,burko}, however, in those cases 
the algorithms were geared to studying gravitational collapse and singularity structure 
in spherical symmetry, and the techniques do not generalize to higher
dimensional systems. Here we present an AMR algorithm for characteristic
evolution that can be applied to scenarios with an arbitrary number of non-trivial spatial dimensions. 
Furthermore,
the algorithm does not place significant restrictions on the discretization
scheme, and hence existing unigrid codes
 (as described, for instance, in \cite{cce,hpgn,luisjcp,gomez,dinverno,bartnik})
can, in principle, be incorporated into the adaptive framework in a
straight-forward manner. To demonstrate the basic algorithm, and show that it
can adapt to dynamical features of a solution, we have implemented a spherically 
symmetric code to solve the Einstein-Klein-Gordon system.

The rest of the paper is organized as follows. In Sec. \ref{sec_eqns} we
describe the coordinate system, set of equations and corresponding discretization scheme
that we will employ in the example problem. We use a coordinate system with a single null direction,
however, the AMR algorithm is most easily presented for double null
coordinates; hence in Sec. \ref{sec_dnull} we describe the adaptive scheme
for double null evolution. In
section \ref{sec_extend} we mention how the basic algorithm is extended to
a coordinate system with a single null coordinate, how additional
non-trivial spacelike dimensions can be handled, mention some problem
specific features such as the numerical dissipation and interpolation 
operators we use, and describe 
a method that can be used to parallelize a characteristic code (with or
without AMR).
In Sec. \ref{sec_results} we present results from our spherically
symmetric code, and give some concluding remarks in Sec. \ref{sec_conclude}.

\section{The spherically symmetric Einstein-Klein-Gordon problem in the characteristic approach}
\label{sec_eqns}
Einstein's equations can be expressed in notational form as
$G_{ab}=8\pi T_{ab}$, where $G_{ab}$ is the Einstein tensor and
$T_{ab}$ the stress energy tensor of the matter distribution. In the particular case of
a scalar field $\Phi$ coupled to GR, $T_{ab}$ results in\cite{hawking}
\begin{equation}
T_{ab} = \nabla_a \Phi \nabla_b \Phi - 1/2 \, g_{ab} \left( \nabla_c \Phi \nabla^c \Phi + m^2 \Phi^2 \right) \, 
\end{equation}
where $g_{ab}$ is the metric tensor, and $m$ is the mass parameter of the scalar field.

We introduce a coordinate system adapted to incoming 
null hypersurfaces in the following way: incoming lightlike hypersurfaces
are labeled with
a parameter $v$, each null ray on a specific hypersurface
is labeled with $(\theta,\phi)$ and $r$ is introduced as a surface area
coordinate (i.e. surfaces at $r={\rm const.}$ have area $4
\pi r^2$). In these coordinates, the metric takes the Bondi-Sachs form\cite{bondi,sachs}
\begin{eqnarray}
   ds^2 & = &  e^{2 \beta} V/r dv^2 + 2e^{2\beta}dvdr +  r^2 d\Omega^2,    \label{eq:bmet_i}
\end{eqnarray}

Note that the choice of incoming null surfaces is merely for convenience; the trivial change
$\beta \rightarrow \beta + i \pi$ takes the line element (\ref{eq:bmet_i}) into the one corresponding to 
outgoing null surfaces (further details can be found in \cite{marsajeff}). The algorithm 
presented here can easily be modified to handle the outgoing case.

In order to express the equations of motion in a simpler form, we introduce the following variables 
\begin{eqnarray}
V &\equiv& -r + g \\
\Psi &=& r \Phi \, .
\end{eqnarray}
The resulting equations (provided by $R_{rr}=8 \pi \Phi_{,r}^2$, $R_{\theta \theta}=8 \pi \Phi_{,\theta}^2$ and $\Box \Phi=0$ 
respectively, where $R_{ab}$ is the Ricci tensor) 
reduce to:
\begin{eqnarray}
\beta_{,r} &=& 2 \pi \, \frac{(r \Psi_{,r}-\Psi)^2}{r^3} \, , \label{beteqn} \\
g_{,r} &=& 1- e^{2 \beta} \left( 1 - 4 \pi m^2 \Psi^2 \right) \label{geqn} \, , \\
2 (\Psi)_{,rv} &=& -\left(1-\frac{g}{r} \right) \Psi_{,rr} + \left(\frac{g}{r} \right)_{,r} \left(\Psi_{,r}-\frac{\Psi}{r} \right)
+ e^{2 \beta} m^2 \Psi \label{phi}
\end{eqnarray}
A properly posed problem requires data to be given on an initial $v=v_o$ hypersurface (which
 consists only of the {\it unconstrained} $\Psi$) and consistent boundary data on an intersecting
surface (which includes $\Psi$, $g$ and $\beta$). The boundary data comprises gauge,
physical and constrained data. For instance, given the value of $\beta$ and $\Psi$ 
on an $r={\rm const.}$ surface the value of $g$ is determined by the remaining Einstein equation
$R_{vv}= 8\pi \Phi_{,v}^2$ (which we call the consistency equation). Next
 we describe the particular setting used in the tests performed 
throughout this work.

\subsection{Coordinate conditions and boundary data}
For our present purposes we choose the outer boundary to coincide with past null infinity (${I}^-$). At ${I}^-$
we fix a Bondi coordinate system, therefore $\beta=0$ (i.e. $v$ represents the affine time of observers
at ${I}^-$). Since past infinity is a null hypersurface, data for $\Psi$ is unconstrained and in fact is intimately
related (just a time derivative difference) to the ``Bondi News'' which is the incoming radiation from 
past null infinity. We provide data for $\Psi$ arbitrarily,
and then determine the value of $g$ using the consistency equation, which on ${I}^-$ reduces to:
\begin{equation}\label{g_init_eqn}
g_{,v} = 8 \pi (\Psi_{,v})^2 \, .
\end{equation}

\subsection{Numerical Implementation}
\begin{figure}
\begin{center}
\psfrag{A}{$f^{n+1}_{i}$}
\psfrag{B}{$f^{n}_{i}$}
\psfrag{C}{$f^{n}_{i+1}$}
\psfrag{D}{$f^{n+1}_{i+1}$}
\psfrag{E}{$f^{n}_{i-1}$}
\psfrag{F}{$f^{n+1}_{i+2}$}
\psfrag{x}{$x$}
\psfrag{v}{$v$}
\psfrag{x0}{$(0,0)$}
\psfrag{v1}{$(1,V_1)$}
\psfrag{xv0}{$(x,v)=(1,0)$}
\includegraphics[width=12cm,clip=true]{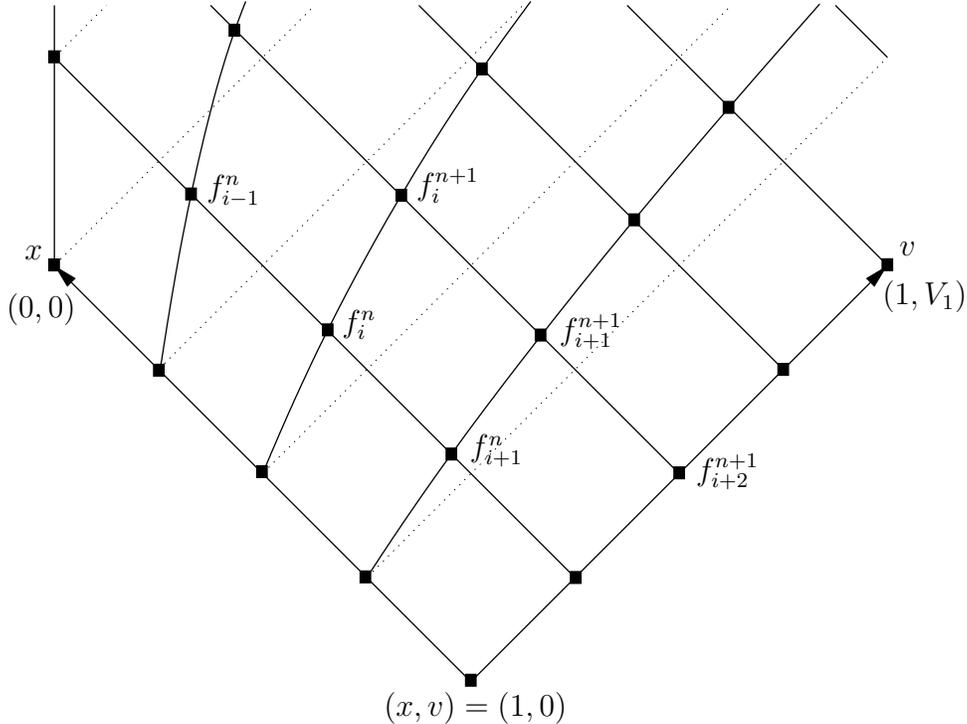}
\end{center}
\caption {A schematic representation of the $(x,v)$ coordinate
system (\ref{eq:bmet_i}) and discretization scheme on a spactime 
diagram, where null directions are at 45 degree angles relative to 
the vertical, and timelike (spacelike) 
curves have tangent vectors less than (greater than) 45 degrees. 
The coordinate lines $v={\rm const.}$ are ingoing null curves,
and $x={\rm const.}$ ($<1$) are timelike curves of constant areal 
radius.
For reference, outgoing null curves (thinner dotted lines) are also
shown on the plot. Note that $x={\rm const.}$ becomes null in the
limit $x\rightarrow 1$, corresponding to past null infinity.
The discretization of a variable $f$ is also shown on the figure---the
points labelled correspond to those that need to be provided (in general)
as ``initial data'' to solve for the unknown $f_{i}^{n+1}$ at 
the interior point ($x_{i},v^{n+1}$).}
\label{bblock_xv_new}
\end{figure}

We discretize equations (\ref{beteqn},\ref{geqn},\ref{phi},\ref{g_init_eqn}) using second order 
finite difference (FD) techniques. In order to include ${I}^-$ in our computational grid, we introduce
a compactified radial coordinate $x=r/(1+r)$ (hence $r \in [0,\infty)$) \cite{gomez} . 
This coordinate is uniform in the domain $[0,1]$
and is discretized by $x_i = (i-1) \Delta x$, with $i=1,2,...,N_x$; similarly, $v$
is discretized so that $v^n=(n-1)\Delta v$, with $n=1,2,...,N_v$. As is customary, we label a grid
function $f$ at coordinate location ($x_i,v^n$) by $f^n_i$---see Fig. \ref{bblock_xv_new} below
for a schematic representation of the coordinate system and discretization.
The boundary data is specified along the pair of intersecting null surfaces $x=1$ (${I}^-$)
and $v=0$. Integration of the evolution
and hypersurface equations (\ref{beteqn}-\ref{phi}) then proceeds via a sequence
of radial integrations (inwards, from $x=1-\Delta x$ to $x=0$) along each of the null 
surfaces from $v=\Delta v$ to $v=V_1$.
Specifically, the discretized hypersurface equations read:

\begin{eqnarray}
\beta^{n+1}_i &=& \beta^{n+1}_{i+1} - 2 \, \pi \, \Delta x \, \frac{(1-x_c)}{x_c^3} \, 
                  \left( x_c (1-x_c) \Psi_{,x}|_o - \Psi|_o \right )^2
                  \label{beta_eqn} \, ,\\
g^{n+1}_i &=& g^{n+1}_{i+1} - \Delta x \, (1-x_c)^2 \, ( 1-e^{2 \beta|_o} (1-4\pi m^2 \Psi|_o^2))\, \label{g_eqn} \, ,
\end{eqnarray}
where 
\begin{eqnarray}
x_c &=& (x_i + x_{i+1})/2 \, ,\\
\Psi|_o &=& (\Psi^{n+1}_i + \Psi^{n+1}_{i+1})/2 \, ,\\
\Psi{,x}|_o &=& (\Psi^{n+1}_{i+1}-\Psi^{n+1}_{i})/\Delta x \, ;
\end{eqnarray}
and similarly for $\beta|_o$.
The evolution equation for $\Psi$ is evaluated  at the point $(v^n+\Delta v/2,x_i+\Delta x/2)$ 
by including 
the points $(v^{n+1},x_i)$, $(v^{n+1},x_{i+1})$, $(v^{n+1},x_{i+2})$,
$(v^n,x_{i-1})$,$(v^n,x_i)$ and $(v^n,x_{i+1})$:
\begin{eqnarray}
\Psi^{n+1}_i &=& - \Big ( - \Psi^{n+1}_{i+1} \, \left ( x_c + \frac{\Delta v}{2\Delta x} (1-x_c)^2 F_1  \right ) \nonumber \\
             & &  +  \Psi^{n+1}_{i+2} \, \left ( \frac{\Delta v}{4\Delta x} (1-x_c)^2 F_1 \right )
                  + x_c (\Psi^n_{i+1}-\Psi^n_i)  \nonumber \\
	     & & + \frac{\Delta v}{4\Delta x} (1-x_c)^2 F_1 (\Psi^n_{i+1} - 2\Psi^n_{i} + \Psi^n_{i-1}) \nonumber \\ 
             & & - F_1 \Delta x \Delta v \Psi_{,x}|_c  (1-x_c) \nonumber \\
             & & + \frac{\Delta v \Delta x}{2} (1-x_c) \left( (1-x_c)g_{,x}|_c-\frac{g|_c}{x_c} \right ) \, \left ( (1-x_c) \Psi_{,x}|_c - \frac{\Psi|_c}{x_c}
\right ) \nonumber \\
             & & + \frac{x_c \Delta x \Delta v }{2 (1-x_c)^2} e^{2 \beta|_c} m^2 \Psi|_c   \Big ) \nonumber \\
             & & / \left ( x_c + \frac{\Delta v}{4\Delta x} (1-x_c)^2 F_1 \right ) \label{psi_eqn}
\end{eqnarray}
where
\begin{eqnarray}
g|_c &=& (g^n_i + g^{n+1}_{i+1})/2 \\
g_{,x}|_c &=& (g^{n+1}_{i+2}-g^{n+1}_{i+1}+g^n_i-g^n_{i-1})/2/\Delta x \\
F_1 &=& -x_c + g|_c (1-x_c) \, , 
\end{eqnarray}
and analogously for $ \Psi_c $, $\beta|_c$ and $\Psi_{,x}|_c$. For each
integration step (\ref{psi_eqn}) is first used to solve for
$\psi^{n+1}_{i}$, then (\ref{beta_eqn}) is solved for $\beta^{n+1}_{i}$, and
finally we obtain $g^{n+1}_i$ from (\ref{g_eqn}). Note that at the
first radial point in from ${I}^-$ (at $x=1-\Delta x$) there are insufficient
points available to evaluate $g_{,x}|_c$ via the above scheme, and
so there we simply propagate $\Psi$ inwards using
\begin{equation}
\Psi^{n+1}_{N_x-1}=\Psi^{n+1}_{N_x}.
\end{equation}

Regularity conditions at $x=0$ are explicitly enforced for $\Psi$
$\{ \Psi(x=0,v)=0 \}$ and $\beta$ $\{ \beta_x(x=0,v)=0 \}$.
Furthermore we set $\Psi(x=\Delta x,v)$ using fourth order interpolation:
\begin{eqnarray}
\Psi^{n+1}_1 &=& 0, \\ 
\Psi^{n+1}_2 &=& 3 \Psi^{n+1}_3/2 - \Psi^{n+1}_4 + \Psi^{n+1}_5/4, \\
\beta^{n+1}_1 &=& 4 \beta^{n+1}_2/3 - \beta^{n+1}_3/3.
\end{eqnarray}

On the outgoing ($x=1$) and ingoing ($v=0$) initial characteristics,
we freely specify $\Psi$. We set $g(x=1,v=0)=2m_0$, where $m_0$ is
the initial, asymptotic mass of the spacetime. We then integrate 
(\ref{beta_eqn}-\ref{g_eqn}) to obtain $\beta(x,v=0)$ and
$g(x,v=0)$, and integrate (\ref{g_init_eqn}) along $x=1$ (using
a similar discretization to (\ref{g_eqn})) to obtain $g(x=1,v)$.

We employ black hole excision techniques to solve for spacetimes containing
a black hole, and consequently a geometric singularity. 
Here the underlying assumption is that {\it cosmic censorship holds}:
any singularity will be hidden by an event horizon (black hole), and
hence cannot causally influence the region outside the black hole\cite{unruh}. This feature
is exploited by placing an inner boundary inside the event horizon, preventing
the simulation from getting too close to the singularity. One cannot determine the
location of the event horizon prior to solving for the geometry of the entire
spacetime; however, we can use the location of an {\it apparent horizon} to tell
us where to excise, for under reasonable assumptions the apparent horizon
always lies inside the event horizon\cite{wald}. An apparent horizon is defined as the outermost,
closed surface whose outgoing null rays form a non-divergent front (i.e the surface
is `trapped'). The location of the apparent horizon is given by $V=0$ in (\ref{eq:bmet_i}).
In practice, when we detect an apparent
horizon, we {\em excise} the portion of the grid interior to it, though we
leave a small buffer zone between the apparent horizon and actual excision 
surface (which is inside the apparent horizon). We implement excision by extrapolating
all variables, using fourth order extrapolation, to all points interior
to the surface of excision. Thus the ``solution'' we obtain inside the black hole
is not physically meaningful, but the excision does {\em not} adversely affect the 
exterior part of the solution that we are interested in.

\section{AMR in double null coordinates}\label{sec_dnull}
Here we motivate and describe our AMR algorithm for characteristic codes.
The salient features of the algorithm are best demonstrated in a 
double null coordinate system; therefore we first describe the 
algorithm in detail for this case, and then discuss
the modifications for the spacelike-null situation in the
following section.

Our AMR algorithm is modeled after the 
Berger and Oliger (B\&O) algorithm \cite{berger} for hyperbolic, Cauchy problems. 
The B\&O algorithm has several desirable features that we have
used as cornerstones in building the scheme for characteristic codes:
\begin{itemize}
\item the computational domain is decomposed into a {\em grid hierarchy},
      whereby the partial differential equations are discretized using
      {\em identical unigrid} finite difference schemes on each grid
      within the hierarchy.
\item dynamical regridding is performed via local 
      {\em truncation error (TE) estimates}.
\item the recursive evolution algorithm makes efficient use of resources
      in both space {\em and} time, for the grids are always evolved
      with a time step set by the {\em local} spatial discretization scale
      (to satisfy the CFL condition), and not by the smallest scale
      within the problem.
\end{itemize}

It is not possible to directly apply the B\&O algorithm in a double null
coordinate system by (for instance) treating one of the null coordinates
as the ``spatial'' coordinate and the other as ``time'', and then integrating the
``spatial'' surfaces in ``time''. This is because propagation along the null surface 
masquerading as a spatial surface will be instantaneous,
and hence the local TE estimation scheme will not be able to track 
corresponding features of the solution. 
The key to adapting B\&O to characteristic codes is to effectively consider
each null direction as ``time'', and then to simultaneously evolve 
along both.

In the double null evolution algorithm the structure of the grid hierarchy 
goes hand-in-hand with the evolution scheme, so we first describe the hierarchy in detail
before presenting the evolution scheme.

\subsection{AMR grid hierarchy}

For the following discussion we will consider the discretization
of a double null coordinate system $(u,v)$, where $u={\rm const.}$ 
labels an outgoing null curve, and $v={\rm const.}$ an ingoing
null curve. For example, a coordinate transformation of the form
$du=-e^{2\alpha} \left(V dv/r + 2dr\right)$, with $e^{2\alpha(v,r)}$ 
an integrating factor, will bring (\ref{eq:bmet_i}) into the form 
\begin{equation}
ds^2 = -e^{2\xi}dudv + r^2 d\Omega^2,
\end{equation}
where $\xi$ and $r$ are now considered functions of $u$ and $v$.

\begin{figure}
\begin{center}
\psfrag{cgs}{\Large Composite Grid Structure}
\psfrag{du1}{$\Delta u_1$}
\psfrag{du2}{$\Delta u_2$}
\psfrag{du3}{$\Delta u_3$}
\psfrag{du4}{$\Delta u_4$}
\psfrag{dv1}{$\Delta v_1$}
\psfrag{dv2}{$\Delta v_2$}
\psfrag{dv3}{$\Delta v_3$}
\psfrag{dv4}{$\Delta v_4$}
\psfrag{u}{\Large $u$}
\psfrag{v}{\Large $v$}
\psfrag{l1}{\Large Level 1}
\psfrag{l2}{\Large Level 2}
\psfrag{l3}{\Large Level 3}
\psfrag{l4}{\Large Level 4}
\includegraphics[width=14cm,clip=true]{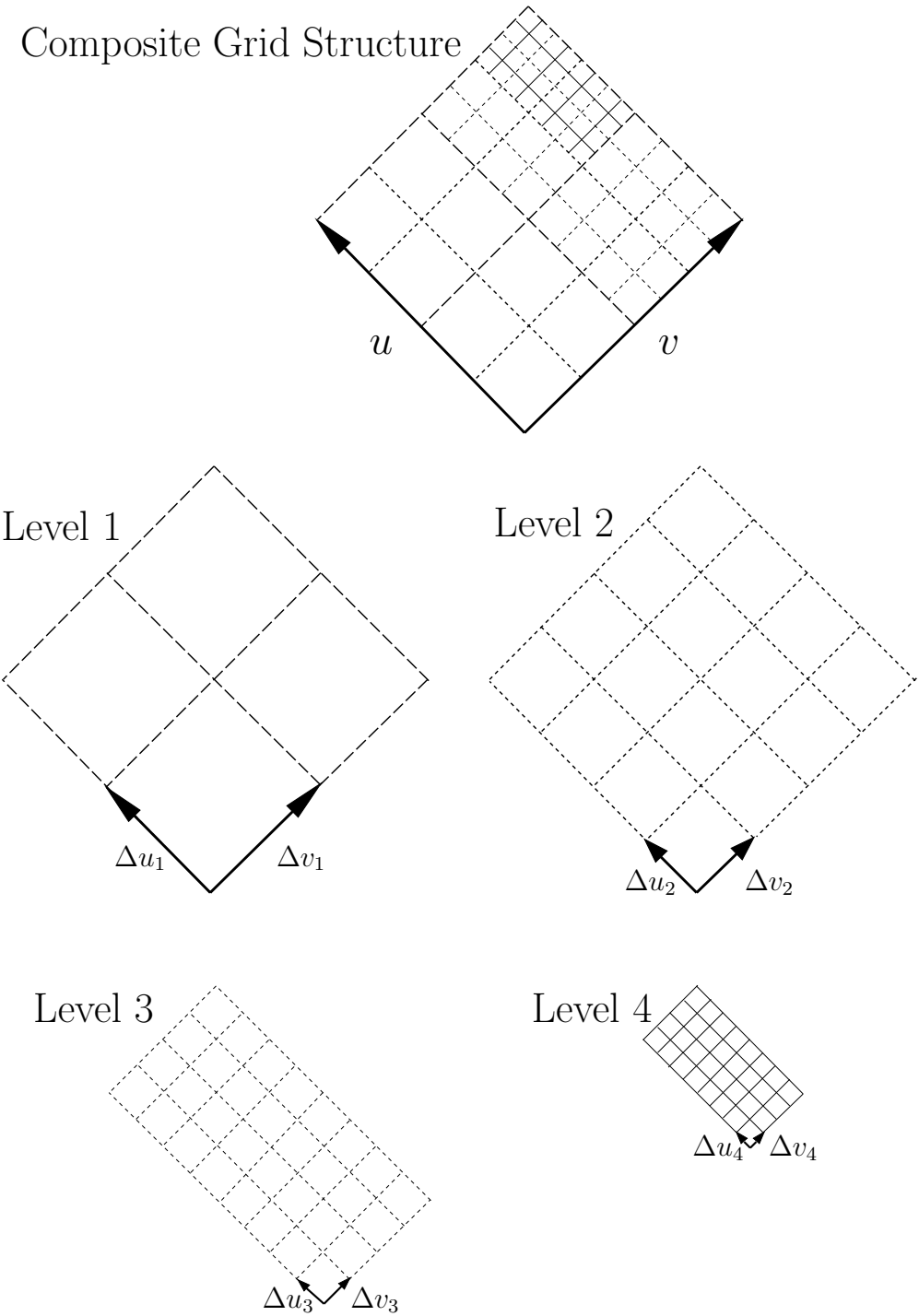}
\end{center}
\caption {An example of a double null grid hierarchy. The 
upper-most figure shows the entire hierarchy, composed of four
levels in this case. Each level is stored
independently of the others; in the lower plots we show the structure
of each of these levels.}
{
\label{sample_mesh}}
\end{figure}

The AMR grid hierarchy (see Fig. \ref{sample_mesh} below for an example)
consists of a sequence of $N$ {\em levels}:
$\ell_1,\ell_2,...,\ell_N$, where all grids at level $m$ are 
discretized on a uniform mesh with  
$[\Delta u_m,\Delta v_m]=[\Delta u_1/\rho^{m-1},\Delta v_1/\rho^{m-1}]$, and $\rho$ is
the {\em refinement ratio}. Therefore, higher levels (larger level numbers) are composed
of finer grids. In this paper we only consider 
$\rho=2$, however the generalization to arbitrary integer 
values of $\rho\ge 2$ is straight-forward (as is the generalization
to different refinement ratios for each level).
All grids at level $m$ are {\em entirely contained} within grids
at level $m-1$. What we mean by this is that the coordinate region
spanned by the union of grids at level $m$ is a subset of the
coordinate region of the union of grids at level $m-1$. 
Furthermore, we require that the meshes of all levels be
aligned such that any point $(u^i,v^j)$ in a grid at level $m-1$
within the region of overlap between 
levels $m$ and $m-1$ (a {\em parent point}) be coincident with 
a point on a grid at level $m$ (a {\em child point}).
This alignment of grids between levels is essential for the recursive
evolution algorithm (and useful for truncation error estimation),
as will be explained in the next section. As an aside, note that the characteristic
AMR algorithm could still allow for rotation of child grids in higher
dimensional ($>2$) simulations in the same sense as the original B\&O 
algorithm, however, here the rotation would be within a subspace
orthogonal to $u={\rm const.}, v={\rm const.}$. In other words,
we only require rigid alignment of the two ``time'' coordinates $u$ and
$v$. 

\subsection{Evolution Scheme}
\begin{figure}
\begin{center}
\psfrag{A}{$A$}
\psfrag{B}{$B$}
\psfrag{C}{$C$}
\psfrag{D}{$D$}
\psfrag{n}{$n$}
\psfrag{s}{$s$}
\psfrag{e}{$e$}
\psfrag{w}{$w$}
\includegraphics[width=8cm,clip=true]{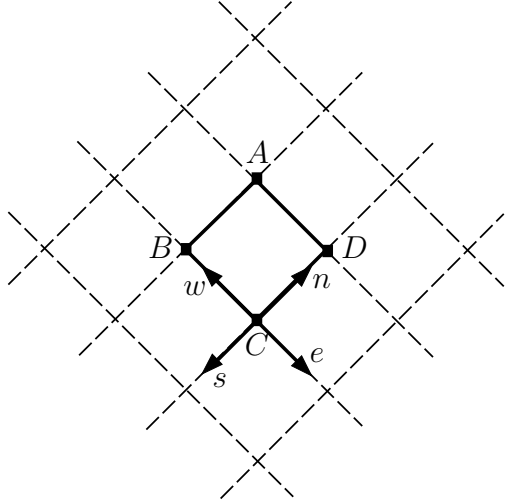}
\end{center}
\caption {A fundamental unit cell of the double null evolution scheme.
 The unit cell consists of the four points
$A,B,C$ and $D$. In the plot we also schematically show part of the data 
structure we use to represent the hierarchy, namely, a set of 
point-structures (or points for short), linked together to form the mesh. 
Each point contains {\em north} ($n$), {\em south} ($s$), {\em east} ($e$) and 
{\em west} ($w$) links to
adjacent points at the same level, as shown for point $C$. In addition, certain
points will have {\em parent} and/or {\em child} links to
corresponding points in the parent and/or child levels (not shown in the
figure).}
\label{bblock}
\end{figure}

Evolution of PDEs discretized on the kind of grid hierarchy just
described proceeds by recursively evolving a particular
sequence of unigrid {\em unit cells}, from the coarsest
to finest levels, and then propagating the solution obtained 
on the finer levels back to the coarser ones.
The unit cell for a typical double null 
discretization scheme is shown in Fig. \ref{bblock}.
For a concrete example, consider the spherically symmetric
wave equation on a flat background (whose line element is
$ds^2=-dudv+r^2 d\Omega^2$, where $r=(v-u)/2$):
\begin{eqnarray}\label{wave_fs}
\Box{\phi}=0 \rightarrow 
\phi_{,uv}+\frac{1}{2r}\left(\phi_{,u}-\phi_{,v}\right)=0
\end{eqnarray}
A second order accurate finite difference version of (\ref{wave_fs}), 
discretized on the unit cell of Fig. \ref{bblock}, is\cite{iww}
\begin{eqnarray}\label{wave_fs_dis}
\frac{\phi_A+\phi_C-\phi_B-\phi_D}{\Delta u\Delta v} + & &  \nonumber \\
\frac{1}{2r}\left( \frac{\phi_A+\phi_B-\phi_C-\phi_D}{2\Delta u}
                   -\frac{\phi_A+\phi_D-\phi_B-\phi_C}{2\Delta v} \right)
&=& 0.
\end{eqnarray}
Initial data for $\phi$ must be specified on the initial ingoing
and outgoing characteristics, which amounts to specifying
$\phi$ at points $B,C$ and $D$.
Evolution to point $A$ then proceeds by solving (\ref{wave_fs_dis})
for $\phi_A$. Evolution of $\phi$ over an entire uniform
mesh of cells is then trivial (ignoring boundary conditions)---the 
unit cell evolution scheme is repeatedly applied, 
in arbitrary order, to all cells where past values 
(corresponding to points $A,C$ and $D$) of $\phi$ are known, until all
points in the grid are solved for.

\begin{figure}
\begin{center}
\includegraphics[width=12cm,clip=true]{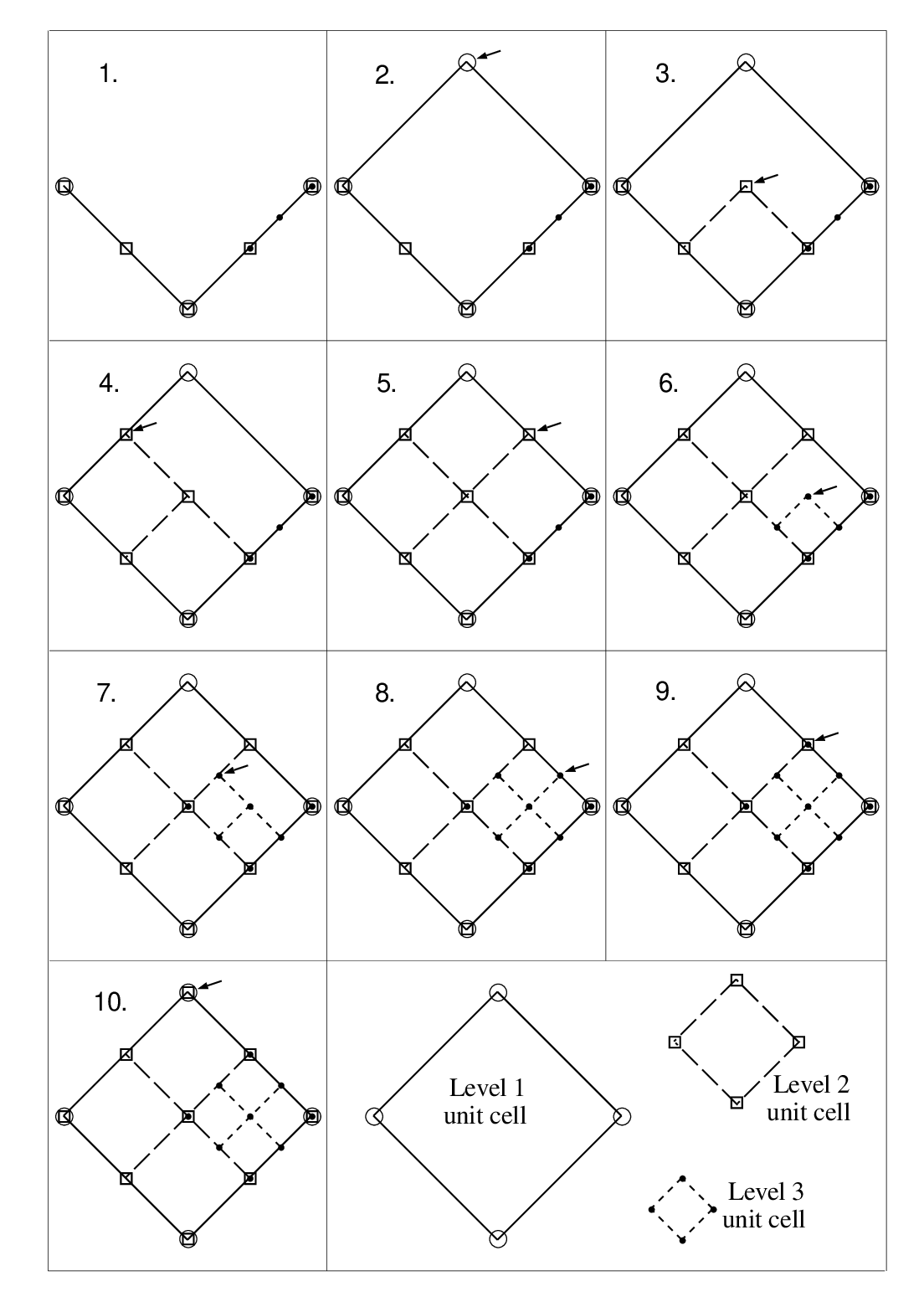}
\end{center}
\caption {Steps in a sample, 3-level evolution.
Frame 1 shows the initial hierarchy. The subsequent frames
show how the hierarchy is recursively created in a single coarse grid
(at level 1) evolution step. An arrow in a frame indicates which grid point 
is being updated during that step. Frames 6 and 7 show the only steps in this
example where data needs to be interpolated from a parent level
(at the point to the ``south'' of the arrowed-point in each case). 
After step 9, we assume that level 3 is unrefined, and hence the final 
evolution step depicted in frame 10 takes place at level 2.
Note that in the algorithm described in the paper, unrefinement (and
equivalently refinement in frame 6) would {\em not} occur here; using
a self-shadow hierarchy, refinement/unrefinement of level 3 can
only occur at points where level 2 and 1 are in sync. For brevity
we ignore this aspect of the algorithm here.
\label{sample_evo}}
\end{figure}

The extension of this unigrid evolution scheme to an adaptive
hierarchy is for the most part straightforward, and follows
the B\&O scheme rather closely in spirit.
The evolution algorithm consists of a particular sequence of
single, unit cell evolution steps. The order in which
cells are traversed over the hierarchy is dictated by {\it causality}, in 
that we can only evolve to point $A$ if
the ``initial/boundary data'' at points $B,C$ and $D$ are known. However, notice in 
Fig. \ref{sample_mesh} (and below in the first frame of Fig. \ref{sample_evo},
depicting a sample evolution) that this initial data
will {\em not} be available on a finer level $m>1$ with initial surfaces 
$u=U_m$ or $v=V_{m}$ that are interior to the computational domain 
boundaries, i.e. where $U_m > U_1$ or $V_m>V_1$ (we assume that at $u=U_1$ and $v=V_1$ the entire
initial grid structure is supplied---see Sec. \ref{sec_extend} 
for a discussion on how the initial hierarchy could be calculated).
The solution to this problem is to always evolve coarser, parent cells
{\em first}, and use the solution obtained on a parent cell to
set initial data, via interpolation, at child cell points bounding
a newly refined region. Then, evolution on the set of child cells is performed,
recursively evolving additional finer levels if present. The
 solution obtained at points coincident with
points on the parent level are {\em injected} back to the parent level, 
to maintain a single-valued 
solution where the most accurately known values are stored at all points in the
hierarchy.

In theory, using interpolation to set interior initial data
of fine regions will not adversely affect the consistency
of the FD approximation to the PDEs
if the hierarchy is generated via
local truncation error (TE) estimates \footnote[1]{We assume that the solution
and  truncation error estimates are sufficiently smooth functions
of $u$ and $v$. However, in principle one can specify non-smooth, though
continuous initial data in $\phi$ on $v=v_0$ and $u=u_0$, together
with the appropriate initial hierarchy.}. For then, prior to the surface
$u=U_\ell$ (or $v=V_\ell$) when a new fine level $\ell$ is added, the local TE 
on level $\ell-1$ will have the same order of magnitude as the maximum allowed TE, and hence
if a sufficiently high order of interpolation is used to initialize
the fields at $u=U_\ell$ ($v=V_\ell$) the solution error
on level $\ell$ will differ from that on level $\ell-1$ by an amount
less that the local TE there.
In practice, interpolation often introduces high frequency solution
components (``noise'') that produce a significant amount of error
when propagated away from the refinement boundaries during subsequent
evolution. Adding numerical dissipation to the FD scheme can
significantly reduce this noise, as can the choice of interpolation
operator and the frequency of regridding. These issues will be discussed
in more detail later on in this section, and in Sec. \ref{custom_ops}.

The rule that we always evolve a parent cell {\it before} any child cells
is naturally implemented via a recursive subroutine, which is
summarized in pseudo-code in Fig. \ref{pseudo_code} below. 
The steps taken in a sample, 3-level evolution is depicted in 
Fig. \ref{sample_evo}.
One of the differences of the characteristic
AMR algorithm, compared to the B\&O algorithm, is that a slightly
more complicated data structure is needed to efficiently represent
the dynamical hierarchy. In B\&O, the grid hierarchy is calculated 
over the entire spatial hypersurface at a given time, and
hence the grids at a given level can efficiently be stored
as a list of one dimensional arrays (in a 1+1 D simulation). 
In the characteristic algorithm, the
structure of the hierarchy is revealed point-by-point,
simultaneously in the $u$ and $v$ directions as one evolves,
and hence one cannot pre-allocate similar one dimensional arrays.
We have chosen to use a data structure 
where a point $(u^i,v^j)$, at some level $\ell$, is the fundamental unit of data.
The mesh is then constructed by linking together adjacent points at
the same level with {\em north($n$), south($s$), east($e$)} and 
{\em west($w$)} pointers as depicted in Fig. \ref{bblock},
and linking points at the same coordinate location in levels
$\ell-1$ and $\ell+1$ via {\em parent} and {\em child} 
pointers respectively. In the pseudo-code in Fig. \ref{pseudo_code}
we have used {\bf C} programming language notation to represent links;
for example, referring to Fig. \ref{bblock}, ${\rm A=B\rightarrow n}$, 
and ${\rm B=A\rightarrow s}$.  In the following paragraphs we will discuss key lines of the
pseudo-code in detail.

\begin{figure}
\input{pseudo_code_jcp.tex}
\caption {A pseudo-code description of the adaptive evolution algorithm.
The function {\em evolve\_unit\_cell(C)} recursively evolves the 
PDEs on all points in the grid hierarchy at the level of point $C$ and higher
one unit cell (of the level of $C$) to the causal future of $C$ (i.e. to
point $A$ in Fig. \ref{bblock}). The function {\em evolve\_hierarchy()}
demonstrates one possible sequence of $evolve\_unit\_cell()$ calls that can
be used to solve the PDEs over the entire hierarchy (we arbitrarily
decided to evolve in $u$ first, then $v$).
\label{pseudo_code}}
\end{figure}

The function {\em evolve\_unit\_cell(C)} listed in Fig. \ref{pseudo_code}
takes, if possible, a single evolution step to the causal future of 
the point $C$, at the level of point $C$. Prior to solving the PDEs
in line $16$, the hierarchy is extended to points $A,B$ and $D$ if
necessary (lines $6-14$). As discussed in the preceding paragraphs, when the
program execution reaches line $6$, points $B$ and $D$
will always exist on the coarsest level of the hierarchy, and at
{\em interior} points of all levels; only on the boundaries of a refined
region might one need to create these points and initialize them via
interpolation from the parent grid. Once the equations have been solved
at point $A$, if the TE at $C$ is greater
than the maximum allowed, we recursively evolve the four unit cells
of the child level that occupy the same region as the unit cell of
point $C$ (lines $18-23$)\footnote[1]{In general, a refinement ratio
of $n:1$ will require a sequence of $n^2$ evolution steps on the
child level at this stage in the algorithm.}.
Note that because of the manner in which
we refine (lines $31-34$), $C$ will always have a child point in the hierarchy
if the TE at $C$ {\em is} greater than the maximum allowed value.
After all the child-levels have been evolved to point $A$, the solution
obtained there on the finest level is {\em injected} back to the
current level (line $30$); i.e. we replace the variables at $A$ with 
those at the child of $A$ (the recursion guarantees that the
values stored at the child of $A$ will have come from the solution
obtained on the finest level of the hierarchy containing $A$).

We compute the truncation error estimate in lines $25-27$ using a
{\em self-shadow hierarchy} technique \cite{fp}, which is a variant of
a {\em shadow hierarchy}. In the traditional B\&O algorithm,
when truncation error estimates are needed at regridding time
two copies of the subset of the hierarchy over-which regridding will
be performed is made; the first is an identical copy of the hierarchy,
while the second is a $2:1$ coarsened version of the first. Then a single evolution
step is taken on the coarsened copy, and two evolution steps (with
the same Courant factor) are taken on the fine copy, i.e. the
copies are evolved to the same coordinate time. The TE
is then computed as a point-wise norm of the difference between the solutions 
obtained on the two copies (which are subsequently deleted). 
A shadow-hierarchy economizes this process by {\em always} evolving
a 2:1 coarsened version (the ``shadow'') of the main hierarchy
in conjunction with the main hierarchy. The solution obtained on the main hierarchy
is then periodically injected into corresponding grids of the shadow
hierarchy. A {\em self-shadow hierarchy} further economizes the truncation
error estimation process by noting that, due to the recursive
nature of the evolution algorithm, information required to compute a TE 
at level $\ell$ is ``naturally'' available prior to the injection step 
from level $\ell-1$ to $\ell$. For at that stage in the evolution process, 
the solution obtained at the common point $A$ on both levels has
been calculated via independent evolution at two discretization
scales, starting from identical ``initial data'' one unit cell to
the past of point $A$ on level $\ell-1$. Thus, metaphorically speaking,
a hierarchy can act as its own shadow at points where there
are at least two levels of refinement. To implement a self-shadow
hierarchy requires that the base level (level $1$) always
be fully refined, and we then define the TE at a point in level $\ell$,
$\ell > 1$, via the difference in the solutions obtained there 
and at the corresponding point in level $\ell-1$, prior to
injection from level $\ell$ to $\ell-1$.

\subsubsection{More on Truncation Error Estimates}

We conclude this section by discussing a few practical details
concerning the computation of the TE. 
The point-wise TE computed using solutions to wave-like finite-difference
equations is 
in general oscillatory in nature, and will tend to go to zero 
at certain points within the computational domain (we will
discuss this in more detail in the next paragraph), even in regions of relatively
high truncation error. We do not want such isolated points of (anomalously)
small TE to cause temporary unrefinement, for experience suggests that
refinement boundaries are often a significant source of unwanted 
high-frequency solution components. Even though we can, to
some degree, eliminate the high-frequency components via dissipation
(see Sec. \ref{sec_extend}) one would like to avoid situations that 
produce ``noise'' as much as possible. Therefore, in practice,
the TE we use to determine whether we refine or unrefine at 
a given point is an average of the point-wise TE taken over several 
cells to the past of the point. Also, note that when using a self-shadow 
hierarchy, we only compute a point-wise TE when point $A$ is in
sync with its parent; we then define the TE at the three points
$A\rightarrow n$, $A\rightarrow w$ and $A\rightarrow n\rightarrow w$ to be identical to that of $A$.

To give a more quantitative description of the nature of the TE,
we will analyze the sample wave equation and discretization
given above (\ref{wave_fs}-\ref{wave_fs_dis}).
Decompose a solution $\phi$ to the finite difference equation 
$\Screll \phi=0$ (\ref{wave_fs_dis}) as
\begin{equation}
\phi=\phi_0 + \phi_e,
\end{equation}
and decompose the difference operator $\Screll$ as
\begin{equation}
\Screll=\Screll_0 + \Screll_e,
\end{equation}
where $\Screll_0=\partial_{uv}+(1/2r)(\partial_u-\partial_v)$ 
is the continuum operator (\ref{wave_fs}),
and $\phi_0$ satisfies the continuum wave equation ($\Screll_0 \phi_0 = 0$).
Hence $\phi_e$ is the truncation error. 
For the discretization in (\ref{wave_fs_dis}),
the operator $\Screll_e$ takes the form
\begin{eqnarray}
\Screll_e&=&\left[\frac{1}{16r}\left(\partial_{uvv}-\frac{\partial_{vvv}}{3}\right)
                +\frac{\partial_{uvvv}}{24}\right] \Delta v^2  \nonumber \\
         & & + \left[-\frac{1}{16r}\left(\partial_{uuv}-\frac{\partial_{uuu}}{3}\right)
                +\frac{\partial_{uuuv}}{24}\right] \Delta v^2 + O(h^4),
                \label{L_e}
\end{eqnarray}
where $h$ denotes either $\Delta u$ or $\Delta v$. Then
\begin{eqnarray}
\Screll \phi &=& 0\nonumber\\
             &=& (\Screll_0 + \Screll_e)(\phi_0 + \phi_e) \nonumber\\
             &=& \Screll_0 \phi_e + \Screll_e \phi_0 + \Screll_e \phi_e \nonumber\\
             &\approx& \Screll_0 \phi_e + \Screll_e \phi_0, \label{te_wave_eq}
\end{eqnarray}
where in the last step we have assumed that the truncation error $\phi_e$ is of 
order $h^2$, and so to leading order we can ignore the term $\Screll_e \phi_e$.
Considering equation (\ref{te_wave_eq}) to be an evolution equation for the truncation 
error $\phi_e$, and assuming that $\phi_0$ is given, we can see that $\phi_e$ satisfies the
continuum wave equation with source term $-\Screll_e \phi_0$:
\begin{equation}\label{te_wave_eq2}
\Screll_0 \phi_e \approx - \Screll_e \phi_0.
\end{equation}
Therefore, from (\ref{L_e}), the leading order part of the
truncation error will be proportional to third and fourth derivatives of $\phi_0$. During
a numerical evolution, if we are in the convergent regime, then a truncation error
estimate computed as described in the preceding paragraphs will be a 
good approximation to the actual truncation error $\phi_e$, and the numerical
$\phi$ will be close to the desired continuum solution $\phi_0$; hence
we would expect the TE estimate to be proportional to derivatives
of $\phi$ as given by (\ref{te_wave_eq2}), 
which in general will exhibit zero-crossings.

\section{Extensions to the basic algorithm}\label{sec_extend}
In this section we briefly describe two extensions to the
double null algorithm introduced in the previous section---first,
to allow for a coordinate system with a single null coordinate, and second,
extensions to higher dimensional systems. We also suggest a technique 
that can be used to parallelize a characteristic code. 

We restrict the discussion on modification for a single null algorithm to the numerical
scheme and coordinate system presented in Sec. \ref{sec_eqns}, though in general 
no significant changes would be needed to alter the algorithm to
use an outgoing instead of ingoing null coordinate, or use different FD stencils.

\subsection{AMR with a single null coordinate}
\begin{figure}
\begin{center}
\psfrag{A}{$A$}
\psfrag{B}{$B$}
\psfrag{C}{$C$}
\psfrag{D}{$D$}
\psfrag{E}{$E$}
\psfrag{F}{$F$}
\psfrag{n}{$n$}
\psfrag{s}{$s$}
\psfrag{e}{$e$}
\psfrag{w}{$w$}
\psfrag{x}{$x$}
\psfrag{v}{$v$}
\psfrag{x0}{$(0,0)$}
\psfrag{v1}{$(1,V_1)$}
\psfrag{xv0}{$(x,v)=(1,0)$}
\includegraphics[width=12cm,clip=true]{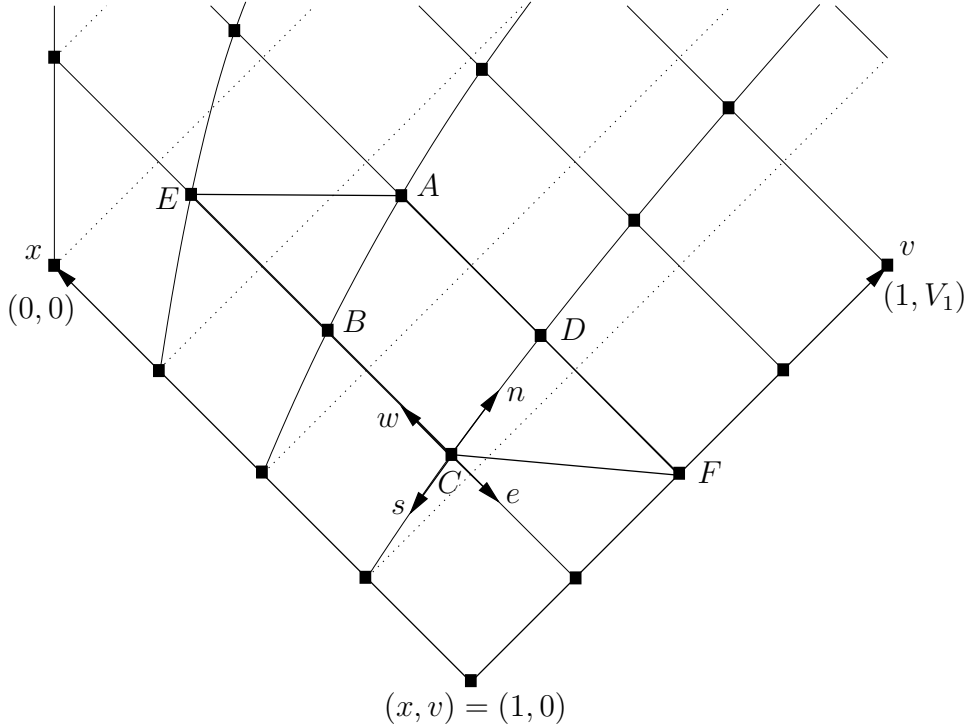}
\end{center}
\caption {The fundamental unit cell of the $(x,v)$ coordinate
system (\ref{eq:bmet_i}) as depicted in Fig. \ref{bblock_xv_new}, and
using the discretization scheme discussed in Sec. \ref{sec_eqns}.
Points $E,B,C,F$ and $D$ hold the ``initial data'' for a single
evolution step that solves for unknowns at point $A$. The same data
structure is used to represent the mesh as with the 
double null scheme (Fig. \ref{bblock}). }
\label{bblock_xv}
\end{figure}

The coordinate system introduced in Sec. \ref{sec_eqns} has a single null coordinate
$v$ and a spacelike coordinate $x$ (which becomes timelike when there are
trapped surfaces). This does not alter the double null algorithm ``in spirit'',
though, as demonstrated in Fig. \ref{bblock_xv} and discussed
further below, the spacelike coordinate
does introduce a preferred direction of integration, and also changes the
shape of the unit cell in a manner that affects the order in which child
cells are recursively traversed during evolution. 
We use the same data structure as before to represent the hierarchy, 
though here $north-south$ links follow lines of 
constant $x$. We also assume that initial data for the evolution is still
specified on a pair of {\em null} surfaces; this is easy
to do in the coordinate system (\ref{eq:bmet_i}) if we use $x=1$ as the
initial surface of integration in $x$, for $x$ becomes
null in the limit $x\rightarrow 1$.

Notice from Fig. \ref{bblock_xv} that because $x={\rm const.}$ is timelike,
causality forces us \footnote[1]{If we want to maintain a relatively simple
time-stepping procedure.} to integrate in $x$, along curves
$v={\rm const.}$, {\em before} taking integration steps in $v$. In other words,
the causal past of the unknown point $A$ at $(x,v)=(x_A,v_A)$, that we want
to solve for during a single integration step, now includes 
regions of spacetime with $x\ge x_A, v\le v_A$ {\em and} $x<x_A$ 
(hence the extension of the unit cell to include point $E$).
Thus, initial data specified along $x=x_0$ and $v=v_0$ 
will not be sufficient to integrate along the sequence of curves
$x=x_0-\Delta x, x=x_0-2\Delta x, ...$; instead
we need to integrate along $v=v_0+\Delta v, v=v_0+2\Delta v, ...$.
Furthermore, note that the size of the integration step $\Delta v$ is now 
subject to a CFL stability condition in that $\Delta v$ must be sufficiently
small so that point $E$ of the numerical stencil is spacelike separated
from $A$.

The change in the causal structure of the coordinate system
also affects the order in which child cells are traversed during
the recursive phase of the evolution algorithm. For the unit
cell depicted in Fig. \ref{bblock_xv}, lines $18-24$ of the double null
algorithm listed in Fig. \ref{pseudo_code} should be modified to the following:
\input{mod_pseudo_code_jcp.tex}
This sequence of child-cell evolution steps ensures that initial data, 
consisting of points $B,C,D,E$ and $F$, is always available when we
integrate to point $A$, at any level
within the hierarchy. Effectively, what we are doing is evolving
all child points that are contained within the coordinate volume of the unit
cell to the past of $A$ {\em before} evolving to the future of $A$. 
The test in line $19$ of the modified algorithm prevents the corresponding
point from being evolved twice in the interior of the grid, which 
otherwise could happen because neighboring unit cells overlap in 
the $x$ direction. The only place where ${\rm C\rightarrow child}$ is evolved
is on the initial $x=X_m$ boundary of a refined 
region or the computational domain (where $X_m=1$). On such a boundary,
if $X_m<1$, we initialize the set of points corresponding to $F$
in Fig. \ref{bblock_xv} via interpolation from parent cells (as is done with the
other points $B,C,D$ and $E$ on the refinement boundary). At $X_m=1$,
we effectively initialize $\Psi$ at $F$ via extrapolation of the initial 
data on $x=1$ to $x=1+\Delta x$ via $\Psi_{,x}=0$ there
(the evolution equations for $\beta$ and $g$ are first order in $x$,
and do not require initial data at $F$).

\subsubsection{Initial hierarchy construction and problem dependent options} \label{custom_ops}
Here we briefly describe some features of our one dimensional code
that are of relevance to a general AMR algorithm, including
the interpolation and dissipation operators we use, though the particular 
implementation of these features may be problem dependent.

We have decided to use the function $\Psi_{,x}$ to compute truncation 
error (TE) estimates. This function behaves adequately in tracking regions of 
high TE in the situations we have looked at, {\em except} 
for incoming waves from ${I}^-$ in the
vicinity of ${I}^-$. The reason why $\Psi_{,x}$ (or any function of
$\Psi$) fails to give a good estimator for the TE of the system there is that,
with our choice of coordinates and variables,
incoming (massless) waves from ${I}^-$ are propagated
essentially without error in the vicinity of ${I}^-$. This has not
been a problem for us, as the initial data we specify on ${I}^-$
is always well resolved with a reasonably sized coarse mesh. Therefore,
the initial hierarchy at $x=1$ only consists of two levels---the base
level ($\ell=2$) and its shadow ($\ell=1$). On the other initial characteristic $v=0$, we 
generate the initial hierarchy by iterating the following until
the number of levels stops increasing: we evolve all levels forward
one coarsest step from $v=0$ to $v=\Delta v_1$ (refining as
usual during the evolution), then we reset $v$ to $0$, and reinitialize 
the fields using the hierarchy obtained at $v=\Delta v_1$. We start the
iteration process with the two coarsest levels fully refined.

Some form of numerical dissipation is usually necessary in Cauchy AMR
codes, for otherwise the interpolation at grid boundaries is often a source
of unwanted, high frequency solution components (``noise'').
We have also
found it necessary to add dissipation to the characteristic 
algorithm\footnote[1]{However here the situation is somewhat different,
in that we need to apply dissipation along a ``time'' direction;
in a typical Cauchy AMR codes one only dissipates in space.}. We
do so by using a Kreiss-Oliger style filter \cite{kreissoliger}, 
whereby we modify $\Psi^{n+1}_i$ after it has been solved for in (\ref{psi_eqn})
via:
\begin{equation}
\Psi^{n+1}_i \rightarrow \Psi^{n+1}_i - \frac{\epsilon}{16} 
\left(\Psi^{n}_{i-2} - 4 \Psi^{n}_{i-1} + 6 \Psi^{n}_{i} - 4 \Psi^{n}_{i+1} + \Psi^{n}_{i+2}\right),
\end{equation}
where $\epsilon$ should be between $0$ and $1$ for stability; we typically use
$\epsilon=0.1$ in the AMR code. This form of dissipation is adequate
in some situations, though is not always effective in reducing noise when a 
new fine level is introduced. Furthermore, even though increasing
$\epsilon$ beyond $\approx0.2$ does help to further reduce the high-frequency noise,
we have found that it often introduces a small, low-frequency,
ingoing ``tail'' component to $\Psi$ when the scalar field is predominantly outgoing. 
Thus, work still needs to be done to find a more effective form of dissipation 
for characteristic AMR. 
A final comment regarding dissipation: after a new fine level is introduced, 
we typically disable regridding in a small buffer zone next
to the boundary of the new level; this gives the dissipation some time
to work at eliminating the noise introduced via interpolation,
and prevents a refinement ``cascade''.

We use cubic Lagrange interpolating polynomials to initialize fine level grid
functions at refinement boundaries if a sufficient number of adjacent
points are available on the parent level, otherwise we use linear 
interpolation---Fig. \ref{pseudo_interp} is a pseudo-code description
of our interpolation routine. We have also experimented with linear
interpolation for all child points, though found that this 
produces significantly more noise after refinement. 

\begin{figure}
\input{pseudo_interp.tex}
\caption {A pseudo-code description of the interpolation routine we use
to initialize functions at points on the initial data surfaces
of interior fine levels. We use cubic interpolation if it is possible
to do so given the structure of adjacent points on the parent level,
otherwise we switch to lower order interpolation. For the kinds of
hierarchies generated by the algorithm described here, one of these
conditions will always be matched, hence the ``if'' part of the 
last ``else if'' statement is not necessary.
\label{pseudo_interp}}
\end{figure}

\subsection{Application to higher dimensional systems}
The extension of the AMR algorithm to higher dimensions, in other words to
problems where there is dependence on additional spacelike coordinates $z^i$, $i=1,2,..,d$, is
straight-forward. The $z^i$ coordinates are treated just as the spacelike
coordinates are in the traditional $B\&O$ algorithm. The null AMR evolution
algorithms described in the preceding sections are un-modified, except
that now at each ``point'' of the mesh structure one needs to store a {\em
list} of $d$--dimensional arrays. Thus, the composite grid hierarchy at any
point $(u,v)$ (or $(x,v)$, etc.) within the computational domain will look 
like a $B\&O$ hierarchy for a $d$-dimensional problem. The particular set
of arrays needed at a given point and at a given level in the hierarchy are
determined, as usual, by local truncation error estimates. Consequently,
standard clustering algorithms will be needed to convert the region
of high TE to a set of grids.

\subsection{Parallelization}
Here we briefly mention a scheme that could be used to parallelize
a characteristic code, with AMR or otherwise. For simplicity,
we illustrate the concept with a unigrid double null scheme, though it is
not difficult to generalize it. The idea is to use
a set of $n$ processors as a pipeline, as demonstrated in Fig. \ref{pipeline} below. 
Integration starts on a single node, where processor $1$ 
solves the equations within a region $R_1$ of size $A=\Delta U\Delta V$
to the future of the initial point $(u_0,v_0)$. Afterward, initial data
is available to simultaneously solve the equations on {\em two}
regions $R_2$ and $R_3$, of the same size $A$, to the future of $R_{1}$.
Processor $1$ (arbitrarily) proceeds to solve the equations in region
$R_{2}$, while supplying the relevant initial data to processor $2$
to evolve region $R_{3}$. After processors $1$ and $2$ have solved 
the equations in regions $R_{2}$ and $R_{3}$ respectively, initial data is now
available to simultaneously evolve {\em three} regions of size $A$.
Processor $3$ can now enter the pipeline, 
and the process continues. After the first $n$ steps, the pipeline
will be full, and all nodes will be involved in the computation.

One desirable feature of this algorithm (compared to a typical
parallelization scheme for a Cauchy problem) is that it would 
be possible to stagger the communication---in other words,
all nodes do not need to, and ideally should not, communicate at the 
same time. Furthermore, in a certain sense the communication is one
way, so that when a processor is finished solving one block
of data it can simultaneously send initial data to
the next processor down the line while starting to evolve a new
block. It would also not be difficult to modify the algorithm
to dynamically subdivide the region of space a processor must
solve in a single step, to provide better load balancing among
the nodes (this will be necessary when AMR is used, or 
in a higher dimensional simulation if the total number of
points on grids in the extra spacelike dimensions depend on
where in $(u,v)$ space one is).

\begin{figure}
\begin{center}
\includegraphics[width=12cm,clip=true]{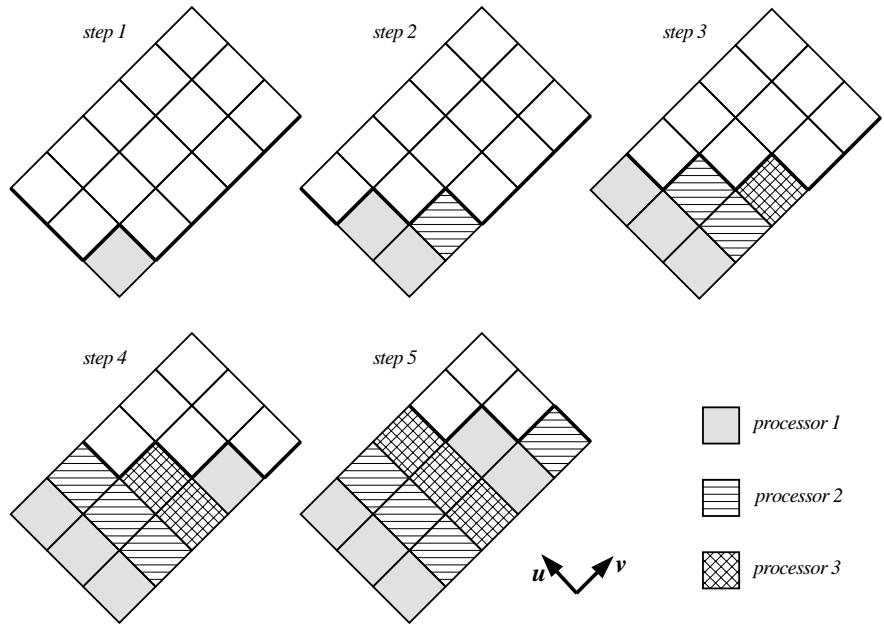}
\end{center}
\caption {An illustration of the technique suggested to parallelize 
a characteristic code. In this example, three processors are available
to simultaneously solve the equations. We subdivide the grid
into blocks of equal area, and assign blocks to individual processors
as shown in the figure. The heavy line drawn on the grid at
step $i$ represents the surface where initial data
will be available at step $i+1$. Before step one, there is only
a single block where sufficient initial data is available
to solve the equations, hence only one processor is active then.
After step one there are two unsolved blocks that can be evolved,
and thus two processors become active; and so on, until the
``pipeline'' of processors is full.
}
{\label{pipeline}}
\end{figure}

\section{Tests \& Results}\label{sec_results}
In this section we present results from two evolutions 
obtained using the 1D characteristic AMR code described in the
previous sections. The first example models a black hole
interacting with an initially outgoing pulse of massive scalar
field energy. The relatively large mass term we use, in conjunction with the
compactified radial $x$ coordinate, causes a wide
range of relevant scales to develop at late times, and we show
that the algorithm is able to track these features with reasonable
accuracy via comparison with results from unigrid evolution. 
The second example shows a near-critical collapse\cite{choptuik2} 
of the massless scalar field. For reasons we will explain below,
our coordinate system is not well suited to studying this kind
of critical phenomena, and hence the example is not very close to criticality.
Nevertheless, we are close enough to demonstrate that the algorithm can
also adapt to the very small length scales that develop from ingoing initial data.

\subsection{Massive scalar field - black hole interaction}\label{sec_massive_results}

\begin{figure}
\begin{center}
\includegraphics[width=15cm,clip=true]{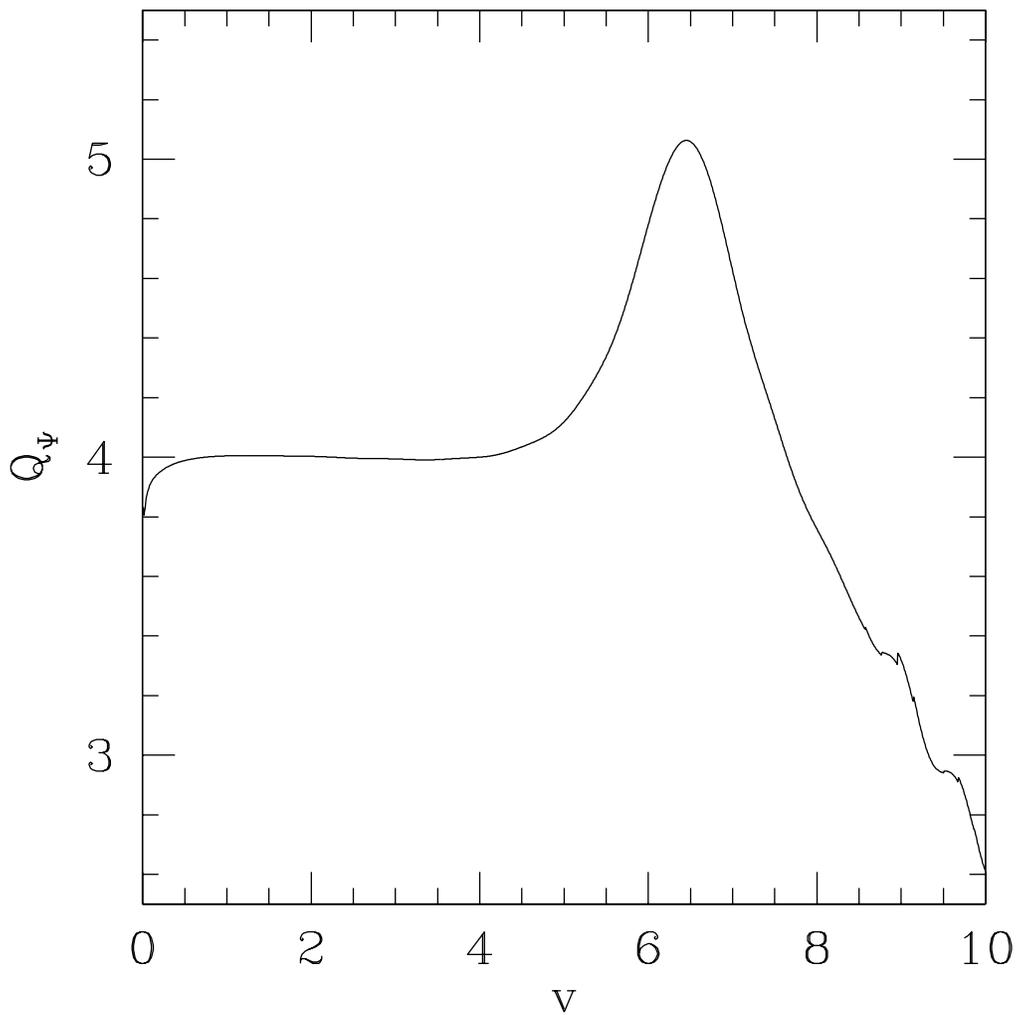}
\end{center}
\caption {The convergence factor $Q_\Psi$ (\ref{q_def}) from
unigrid evolution of the black hole-massive scalar field initial
data.
\label{cvf_l2_psi_123}}
\end{figure}

For the first example, we specify $\Psi$ along $v=0$ as follows:
\begin{eqnarray}
\Psi(v=0,x)&=&A x (1-x/x_1)^4 (1-x/x_2)^4, \ \ \ x_1 < x < x_2,\nonumber\\
           &=&0 \ \ \ {\rm elsewhere},
\end{eqnarray}
choose $x_1=0.15$, $x_2=0.25$, $A=5\rm{x}10^{3}$ and the mass
parameter $m=5$. We set $g(x=1,v=0)=0.2$, so that the asymptotic mass
of the spacetime is $0.1$; the initial ($v=0$) black hole mass is 
$\approx 0.063$, indicating that the scalar field contributes $\approx 0.037$ 
to the initial mass of the spacetime. We ran the simulation to $v=10$, by which
time the black hole mass had grown to roughly $0.068$ due to accretion of scalar
field energy. First, to demonstrate the correctness of our solution to the finite
difference equations, in Fig. \ref{cvf_l2_psi_123} we show the convergence
factor for the scalar field $Q_\Psi$ 
\begin{equation}\label{q_def}
Q_\Psi=\frac{||\Psi_{4h}-\Psi_{2h}||_2}{||\Psi_{2h}-\Psi_{h}||_2},
\end{equation}
computed using solutions from 3 different resolutions of {\em unigrid}
simulations. $\Psi_{4h}$ has 1025 points in $x$, $\Psi_{2h}$ 2049, and
$\Psi_{h}$ 4097. The Courant factor is $0.25$ in all cases; i.e.
$\Delta v=0.25\Delta x$. That $Q_\Psi$ is around $4$ for most of the simulation
indicates that we are seeing second order convergence, as expected
(and that $Q_\Psi$ eventually starts to deviate from $4$ is also not 
unexpected---accumulating numerical errors, especially in our 
compactified coordinate system, will eventually cause any finite resolution 
simulation to move away from the convergent regime).
For brevity we do not show convergence factors for the other fields;
they also exhibit second order convergence. To demonstrate the accuracy of the 
adaptive code, we first show a comparson between a solution 
generated with AMR to the finest unigrid solution, and then
present some results from a convergence test. 

\subsubsection{Comparison between AMR and unigrid solutions}\label{sec_massive_comp_results}

\begin{figure}
\begin{center}
\includegraphics[width=15cm,clip=true]{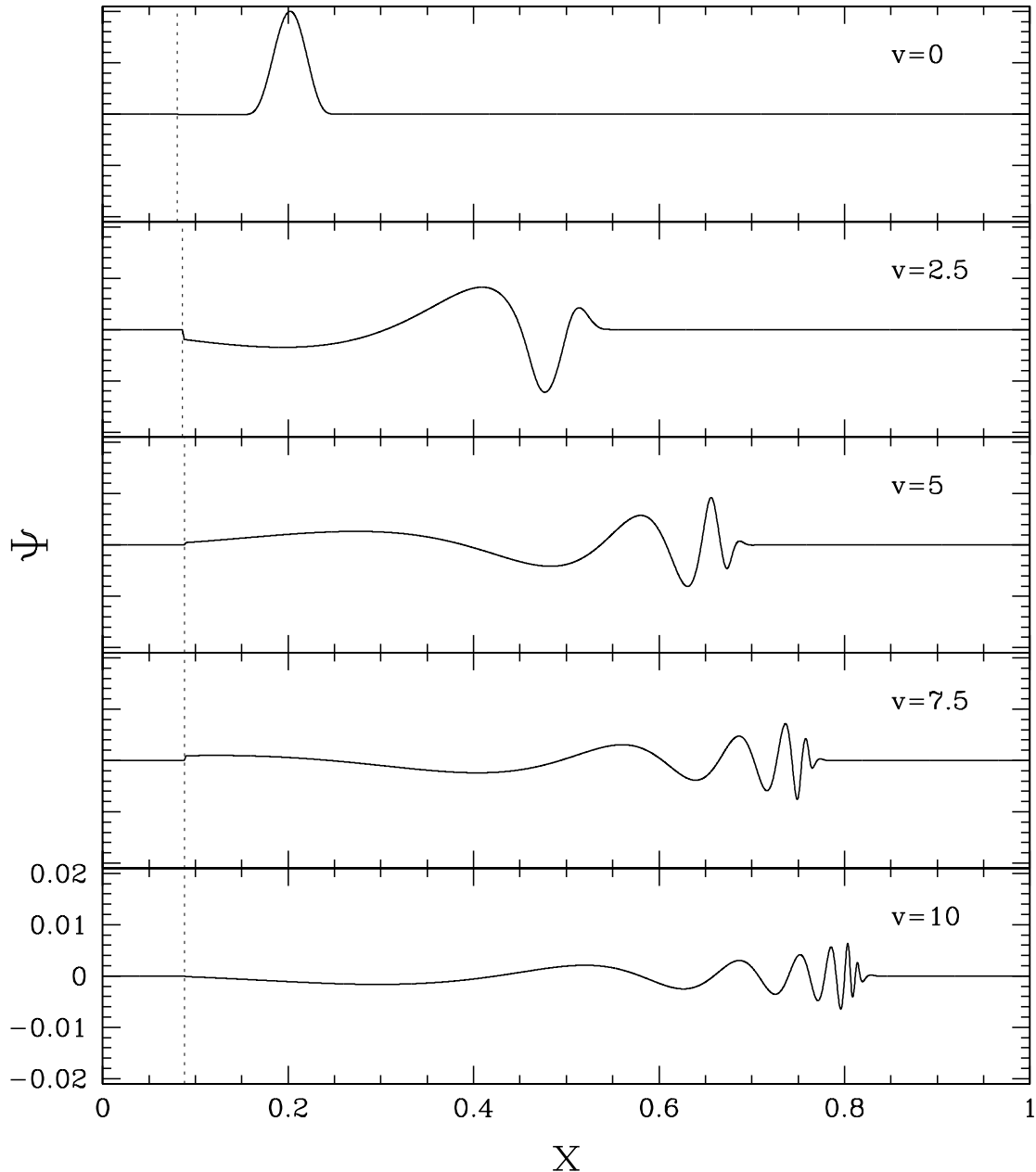}
\end{center}
\caption {$\Psi(x)$ along several $v={\rm const.}$ slices 
of the spacetime, from the adaptive black hole-massive scalar field
simulation discussed in section \ref{sec_massive_comp_results}.
The position of the excision surface is denoted by a
vertical dashed line, and is always kept a distance of $0.025$ in
$x$ inside the apparent horizon; the scalar field is set to zero inside
the excision surface. The scalar field pulse is initially outgoing,
and the higher-frequency components of the field continue to 
travel outward at the speed of light. The non-zero mass term in the 
scalar field equation of motion causes lower frequency components 
of the field to travel at speeds less that $1$; these trail the leading
pulse of the field, and interact more strongly with the black hole. 
See Fig. \ref{uu_ad_lev} below for a plot of the maximum AMR level
along the same slices shown here to see how the AMR algorithm
correctly tracks the outgoing, higher-frequency components of the
solution.
\label{uu_psi_ad}}
\end{figure}

\begin{figure}
\begin{center}
\includegraphics[width=15cm,clip=true]{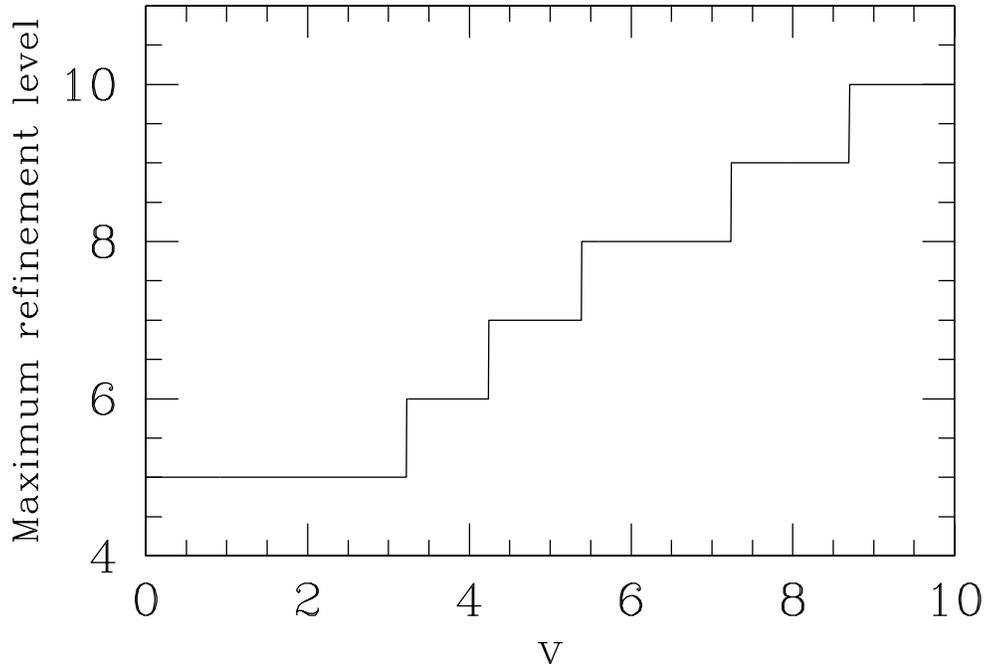}
\end{center}
\caption {Maximum level of the adaptive hierarchy along
$v={\rm const.}$ slices of the spacetime, from the adaptive
black hole-massive scalar field simulation 
discussed in section \ref{sec_massive_comp_results}.
\label{ad_max_lev}}
\end{figure}

For the comparison between the AMR and unigrid results, the parameters
for the AMR run were set so that the base level (2) has $513$ points
(hence the shadow level (1) has $257$ points), and the maximum
allowed TE is such that early on (in $v$) during the evolution 
the finest level is $5$, giving the same resolution locally
as that of the finest unigrid simulation. At late times during the
adaptive run additional levels are introduced
to track the outgoing pulse (whose width shrinks due to
the radial coordinate we use)---see Fig. \ref{uu_psi_ad}
for several snapshots of $\Psi$ along $v={\rm const.}$ surfaces
during evolution, Fig. \ref{uu_ad_lev} for the 
maximum level as a function of $x$ at the same
instants of $v$ shown in Fig. \ref{uu_psi_ad}, and
Fig. \ref{ad_max_lev} for a plot of the maximum
level in the hierarchy over all $x$ as a function of $v$.
To gauge the quality of the adaptive solution, we use the
highest resolution unigrid solution for $\Psi$ as a benchmark,
and compare this solution to the lower resolution unigrid
and adaptive results. Fig. \ref{l2_dpsi_comp} shows
the $\ell_2$ norm (computed along $v={\rm const.}$ slices of the
spacetime) of these differences, and Fig. \ref{psi_comp}
plots $\Psi(x,v=10)$ from the four simulations near the
leading edge of the pulse, for visual comparison. Fig. \ref{l2_dpsi_comp}
demonstrates that early on the adaptive solution gives slightly
worse results than the lower resolution unigrid solution,
which is not too surprising as the AMR solution only covers 
a portion of the computational domain with comparable or higher resolution.
However, at late times the adaptive solution starts to 
outperform the coarser unigrid solutions as the AMR is able to keep the 
narrowing pulse well resolved. 

\subsubsection{AMR convergence test}\label{sec_massive_cvt_results}

\begin{figure}
\begin{center}
\includegraphics[width=15cm,clip=true]{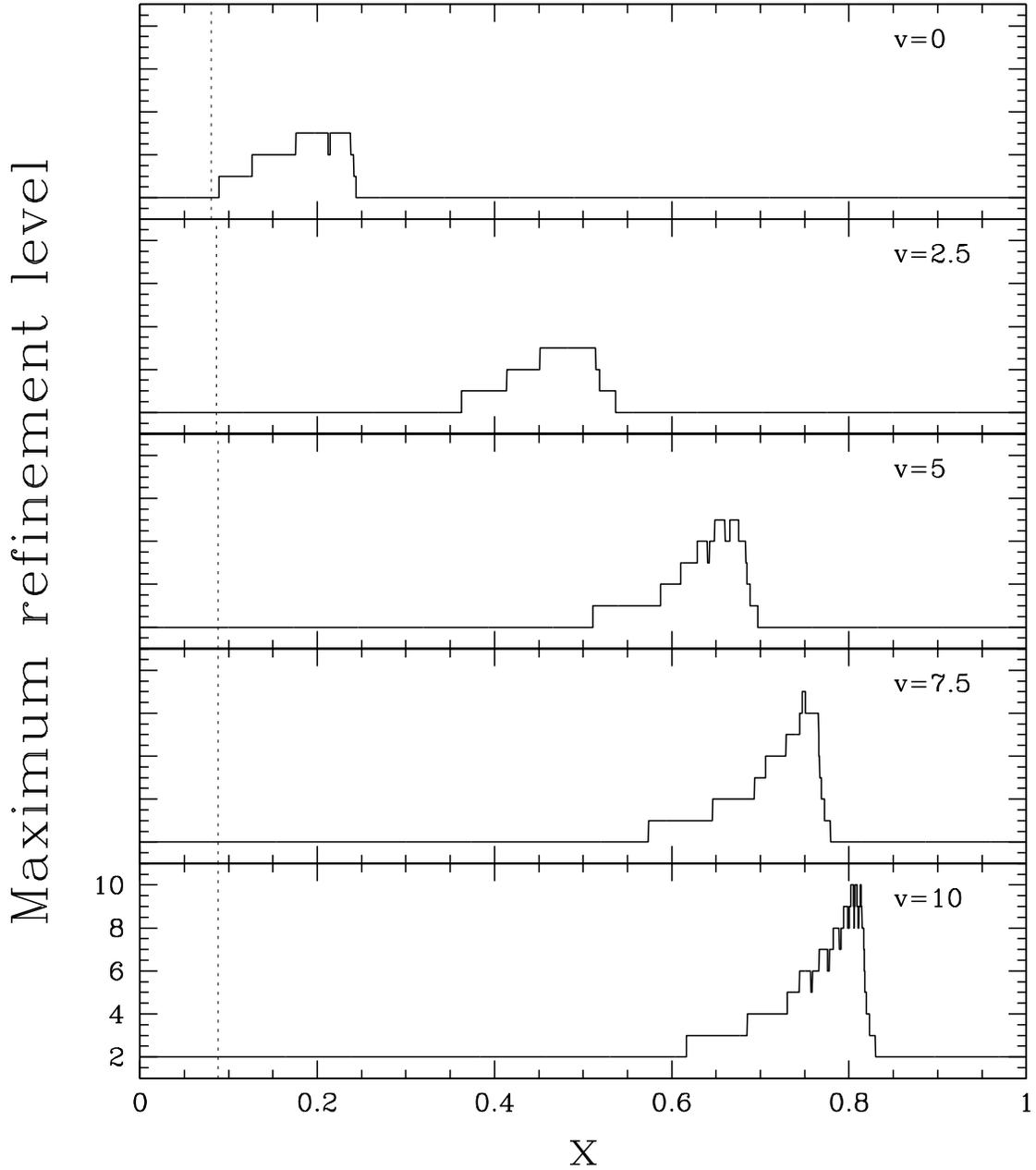}
\end{center}
\caption {Maximum level of the hierarchy from the adaptive
black hole-massive scalar field simulation discussed in section \ref{sec_massive_comp_results}, 
at the same
slices of the computational domain shown in Fig. \ref{uu_psi_ad}.
\label{uu_ad_lev}}
\end{figure}

\begin{figure}
\begin{center}
\includegraphics[width=14.5cm,clip=true]{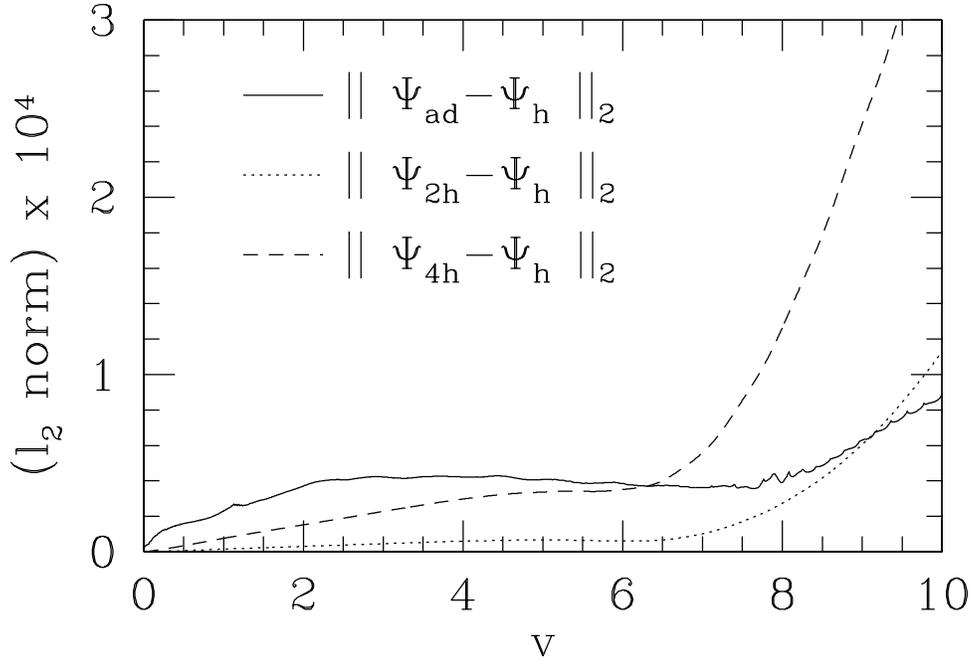}
\end{center}
\caption {Comparison of $\Psi(x,v={\rm const.})$ between
the highest resolution unigrid simulation ($\Psi_h$), and
lower resolutions unigrid simulations ($\Psi_{2h}$ and $\Psi_{4h}$) 
and adaptive simulation ($\Psi_{ad}$) described in 
in section \ref{sec_massive_comp_results}.
The unigrid $h$ ($2h,4h$) simulation
has $4097$ ($2049,1025$) points in $x$, while level
$5$ of the AMR run has the same resolution as the $h$
unigrid run (see Fig. \ref{uu_ad_lev}). This figure shows that
at early times, both the $2h$ and $4h$ unigrid simulations perform
slightly better than the adaptive run, compared to the
$h$ unigrid solution; however at late times the adaptive
code starts to outperform the lower resolution unigrid runs, as
the AMR tracks the ever-narrowing outgoing pulse
(see Fig. \ref{uu_psi_ad}). See Fig. \ref{psi_comp} below for plots
of $\Psi$ from the four solutions at $v=10$.
\label{l2_dpsi_comp}}
\end{figure}

\begin{figure}
\begin{center}
\includegraphics[width=14.5cm,clip=true]{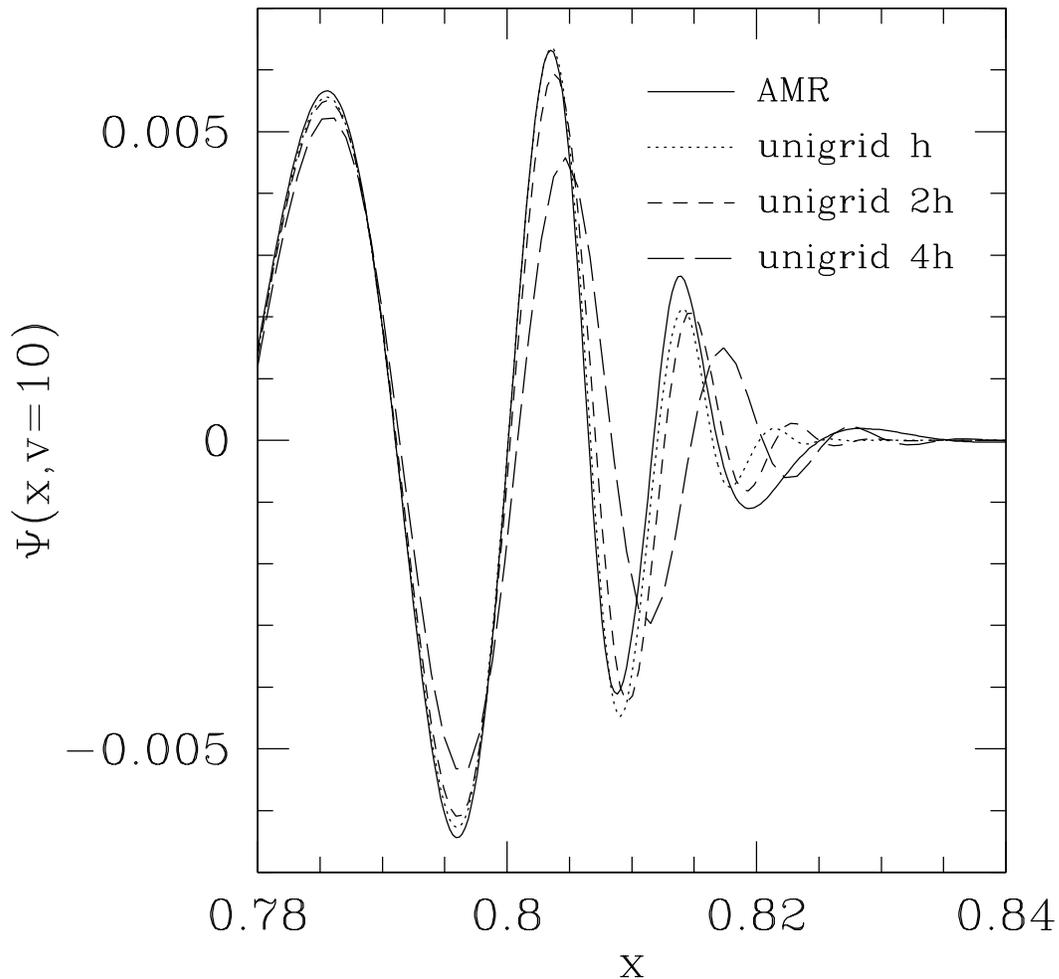}
\end{center}
\caption {$\Psi(x)$ from the adaptive and
three unigrid solutions, at $v=10$ of the
black hole-massive scalar field simulation (section \ref{sec_massive_comp_results}). 
We only show a small range of the $x$ coordinate
domain (compare Fig. \ref{uu_psi_ad}), as 
this region shows the largest disagreement among
the four simulations. The unigrid $h$ ($2h,4h$) simulation
has $4097$ ($2049,1025$) points in $x$, while level 
$5$ of the AMR run has the same resolution as the $h$ 
unigrid run (see the last frame of Fig. \ref{uu_ad_lev}).
\label{psi_comp}}
\end{figure}

For a separate convergence test of the AMR code we ran three
simulations with the same initial data as described above,
varying the base grid resolution and maximum allowed
truncation error to mimic doubling the resolution from one
run to the next. Specifically, the lower resolution simulation
had 257 points in the base level (2) with a maximum TE of
$\tau_{4h}$, the medium resolution run had a base grid of 513 points
and a maximum TE of $\tau_{2h}=\tau_{4h}/4$, while the
higher resolution simulation had a base grid of 1025 points
and a maximum TE of $\tau_{h}=\tau_{4h}/16$. However, due to limited
available computer resources we restricted the maximum depth of the
hierarchy to 7 for these three runs (at this stage our code is not 
memory efficient, and the high resolution run was using 
most of the available memory nearing the end of the simulation).
Note that this
scheme for convergence testing an AMR code will {\em not} produce
identical grid hierarchies that differ only in resolution, because
the truncation error {\em estimate} used in each numerical simulation
will not scale exactly as the leading order part of the actual truncation
error, which decreases by a factor of 4 each time
the mesh spacing is halved for a second order accurate scheme.
Nevertheless,
if the test were to show non-convergent results it would be a clear
indication of problems with the implementation.
Fig. \ref{cvf_ad_l2_psi} below shows the convergence factor
for $\Psi$ (where to calculate $Q_\Psi$ as in (\ref{q_def}) 
we first interpolated the solutions to identical uniform grids).
Early on we get a convergence factor that is closer to first than second
order convergence. The primary reason for this appears to be grid-boundary
``noise'' generated at parent-child interfaces, and as discussed in 
Sec. \ref{custom_ops} our current interpolation/dissipation scheme is not yet very
effective at reducing this noise. Later on in the simulation
the convergence behavior
appears to improve significantly (and become unrealistically high), 
however this is mostly because
we had to limit the maximum level to 7. At late times 
this is not sufficient to maintain the desired
TE (see Fig. \ref{uu_ad_lev} for the level structure generated
by the AMR run described in the previous section, which had
a base resolution equivalent to that of the medium 
resolution simulation here), and the solutions start to drift away from the convergent
regime. In fact, by $v=10$ the lower resolution run had accumulated significant
phase errors in $\Psi$, whereas the medium and high resolution
solutions were still roughly in phase, which gives some explanation for
the anomalously high value of $Q_\Psi$ (\ref{q_def}) in this case.
 
\subsection{Massless scalar field critical collapse}\label{sec_massless_results}

\begin{figure}
\begin{center}
\includegraphics[width=15cm,clip=true]{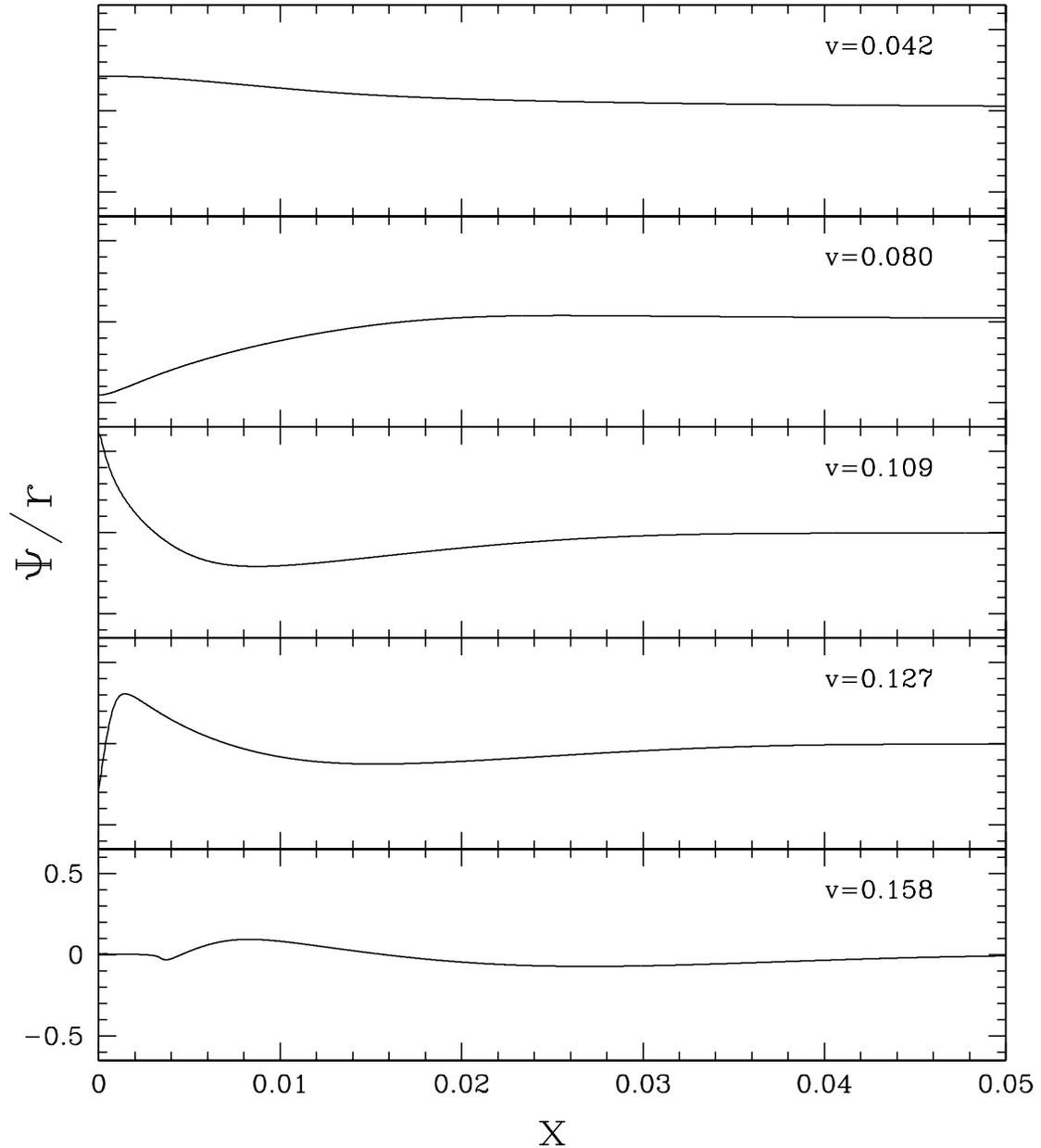}
\end{center}
\caption {$\Psi(x)/r(x)$ along several $v={\rm const.}$ slices 
of the spacetime, from a sub-critical, massless scalar field
evolution (with purely ingoing initial data, i.e. $\Psi(x,v=0)=0$).
The first four plots above correspond to slices where
$\Psi(x)/r(x)$ has attained a local maximum or minimum at the
origin; the first two of these would be present in the 
weak field regime, in the next two we are beginning to see
the first half-echo of the critical solution (one feature
of which is that the central value of the scalar field
oscillates between approximately $\pm 0.6$ during each self-similar
echo). In the last frame the scalar field is dispersing.
\label{uu_c_phi_ad}}
\end{figure}

\begin{figure}
\begin{center}
\includegraphics[width=15cm,clip=true]{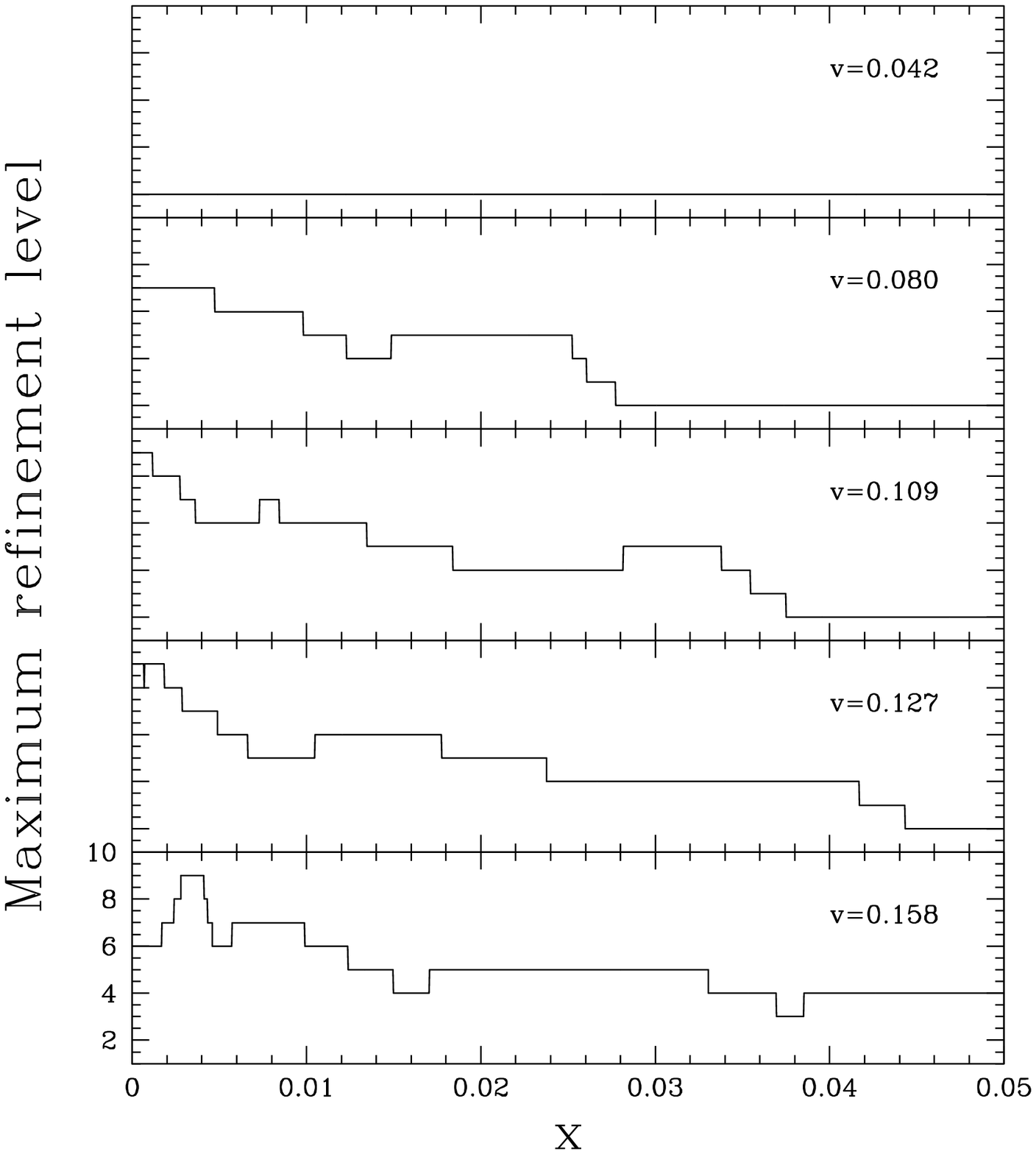}
\end{center}
\caption {The maximum level of the hierarchy along the
slices of the near-critical simulation shown in Fig. \ref{uu_c_phi_ad} 
\label{uu_c_phi_ad_levs}}
\end{figure}

For a second, brief example, we consider the near-critical collapse of
an initially ingoing pulse of the massless scalar field:
\begin{eqnarray}
\Psi(v,x=1)&=&A (1-v/v_1)^4 (1-v/v_2)^4, \ \ \ v_1 < v < v_2,\nonumber\\
           &=&0 \ \ \ {\rm elsewhere},
\end{eqnarray}
and set $v_1=0.01$, $v_2=0.11$ and $g(x=1,v=0)=0$. $A$ is 
tuned (via a bisection search) so that
the collapsing pulse is close to the threshold of black hole formation.
As mentioned before, the coordinates we use are not well suited 
to studying type II critical collapse, where one should be able 
to form arbitrarily small black holes in the super-critical regime.
The reason is that in our coordinate system, a non-zero initial
value $g(x=1,v=0)=2m_0$
describes a spacetime containing a black hole of mass $m_0$ (assuming
$\Psi(x,v=0)=0$).
Truncation error effects in the integration of $g(x=1,v)$, and subsequent
evolution in $x$, causes small errors in $g$ that effectively
behave as if a small black hole (of size proportional to the TE) 
had been present in the initial data. This is not a problem for unigrid
evolution, as the erroneous black hole is typically smaller than the grid
spacing; however during a critical evolution where arbitrarily small scales
unfold, and refinement resolves these scales, this black hole is eventually
revealed. Thus the resolution of the initial data at ${I}^-$ places a 
limit on how close to critical we can evolve\footnote[1]{To seriously 
study critical collapse with this characteristic AMR algorithm one would therefore
need to choose more appropriate coordinates, such as one based on an outgoing
null coordinate for instance \cite{garfinkle2}.}. For this example 
we chose a base (level $2$) resolution of $2049$ in $x$ (with $\Delta v=0.25\Delta x$);
then the smallest black hole we can form is 
on the order of $10^{-4}$, which is roughly $1/2$ the size of a base-level
cell in $x$.
Fig. \ref{uu_c_phi_ad} shows several snapshots of $\Psi/r$ from the evolution of 
the nearest-to-critical sub-critical amplitude we found, and 
Fig. \ref{uu_c_phi_ad_levs} shows the corresponding level 
structure. 

\begin{figure}
\begin{center}
\includegraphics[width=15cm,clip=true]{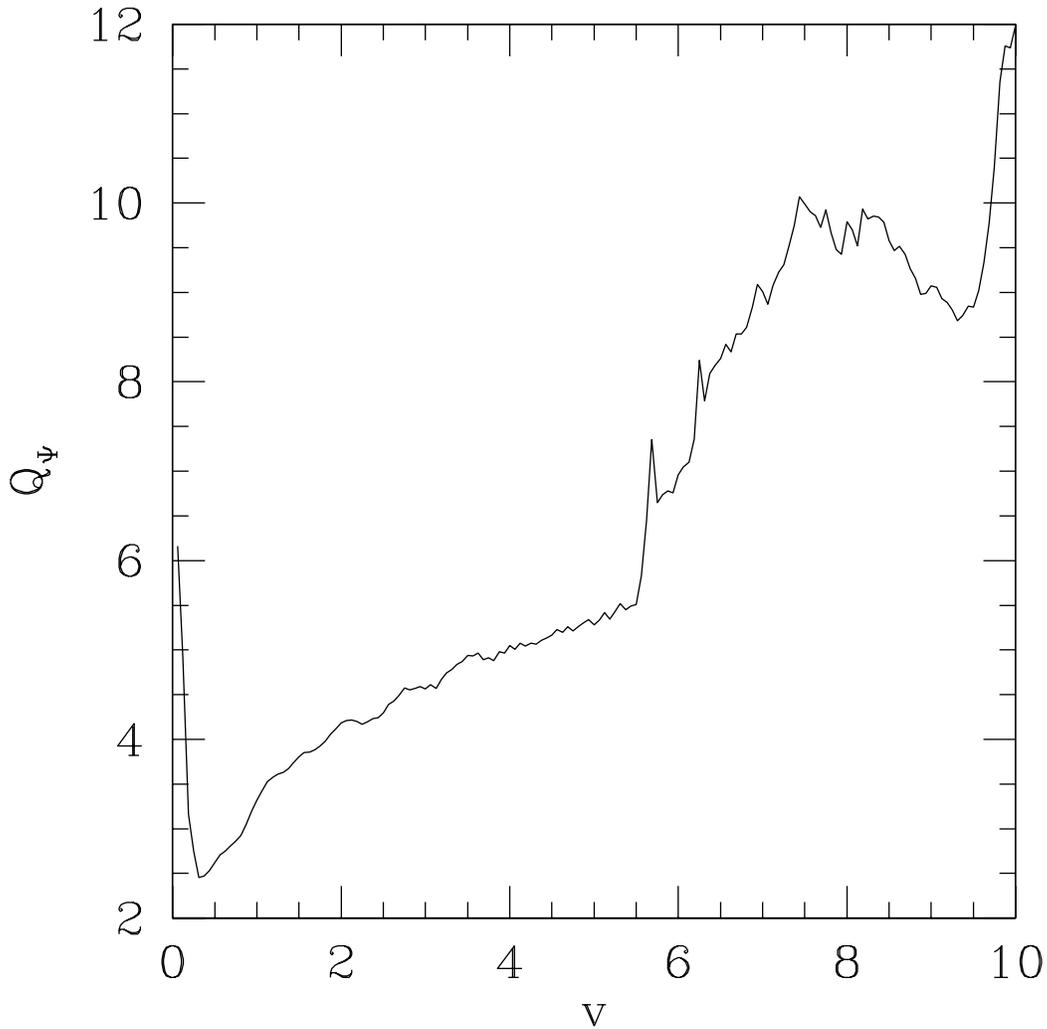}
\end{center}
\caption {The convergence factor $Q_\Psi$ (\ref{q_def}) from
AMR evolutions of the black hole-massive scalar field initial
data. The lack of second order convergence is primarily
due to inadequate dissipation of high-frequency ``noise''
generated at parent-child interfaces, and the
anomalously high convergence factor at later times is due
to a significant phase error developing in the solution
from the lowest resolution run compared to the two higher
resolution runs; see the text in section \ref{sec_massive_cvt_results} 
for further discussion.
\label{cvf_ad_l2_psi}}
\end{figure}

\section{Conclusion}\label{sec_conclude}
In this paper we have introduced a new algorithm that can be used
to add an adaptive mesh refinement framework to a characteristic evolution code.
The algorithm is similar to the Berger and Oliger algorithm for
Cauchy codes, and shares many of its desirable features,
including dynamical regridding via local truncation error estimates,
efficient use of computational resources via the recursive evolution
scheme, and in principle it does not require modifications to
the underlying unigrid finite difference scheme. As discussed in
the introduction, we believe AMR is essential to achieve 
accurate results from simulations in many general relativistic scenarios.
Based upon the early success of this AMR technique with the one dimensional
code presented here, we think it would be well worth the effort 
to apply the method to higher dimensional problems of interest. 

\section{Acknowledgments}
We thank Matthew Choptuik for valuable insights.
FP would like to acknowledge
NSERC, The Izaak Walton Killam Fund, Caltech's Richard Chase Tolman Fund
and NSF PHY-0099568 for financial support.
Computations were performed on the vn.physics.ubc.ca Beowulf cluster
(supported by CFI and BCKDF)
L.L. thanks PIMS and CITA for partial finantial support and the California Institute 
of Technology for hospitality where the last stages of this work were completed.


\bibliography{refer}
\bibliographystyle{elsart-num}

\end{document}

%% file: pseudo_code_jcp.tex
{\small
\renewcommand{\baselinestretch}{0.75}
\begin{verbatim}
 1:   logical function evolve_unit_cell(point C)
 2:      if (can_take_step(C)==false) then 
 3:         return(false);
 4:      end if
 5:   
 6:      if (point B=C->w does not exist) then
 7:         create B and link it into the surrounding mesh;
 8:         initialize variables at B via interpolation from parent points;
 9:      end if
10:      if (point D=C->n does not exist) then
11:         create D and link it into the surrounding mesh;
12:         initialize variables at D via interpolation from parent points;
13:      end if
14:      create point A=D->w(=B->n) and link it into the surrounding mesh;
15:   
16:      solve for the variables at A via the evolution equations;
17:   
18:      if (TE(C)>maximum_TE) then 
19:         evolve_unit_cell(C->child); 
20:         evolve_unit_cell(C->child->w);
21:         evolve_unit_cell(C->child->n);
22:         evolve_unit_cell(C->child->n->w);
23:      end if
24:   
25:      if (A->parent exists) then
26:         compute_TE(A);
27:      end if
28:   
29:      if (A->child exists) then
30:         inject variables from A->child to A;
31:      else if (TE(A) > maximum_TE) then
32:         create point A->child;
33:         initialize variables of A->child with the corresponding values at A;
34:      end if
35:   
36:      (one can safely delete points to the causal past of C here)
37:   
38:      return(true);
39:   end of function evolve_unit_cell;
40:
41:   function evolve_hierarchy()
42:      Nu1=number of points in the u-direction in the base level(1);
43:      Nv1=number of points in the v-direction in the base level;
44:      
45:      create the hierarchy corresponding to the initial surfaces
46:         u=u0 and v=v0, and initialize the variables there;
47:
48:      set point Ci = grid point at (u0,v0) on the base level;
49:      for i=1 to (Nv1-1) do
50:         Cj=Ci;
51:         for j=1 to (Nu1-1) do
52:            evolve_unit_cell(Cj);
53:            Cj=Cj->w;
54:         end do
55:         Ci=Ci->n;
56:      end do
57:   end of function evolve_hierarchy;
\end{verbatim} }

%% file: mod_pseudo_code_jcp.tex
{\small 
\begin{verbatim}
18:      if (TRE(C)>maximum_TRE) then 
19:         if (C->child has not been evolved) then evolve_unit_cell(C->child); 
20:         evolve_unit_cell(C->child->w);
21:         evolve_unit_cell(C->child->w->w);
22:         evolve_unit_cell(C->child->n);
23:         evolve_unit_cell(C->child->n->w);
24:      end if
\end{verbatim} }

%% file: pseudo_interp.tex
{\small
\renewcommand{\baselinestretch}{0.75}
\begin{verbatim}
subroutine interpolate (grid function f, point C)
   if (C->parent exists) then

      f[C] = f[C->parent]; (straight copy)

   else if ( [C->s,P2=C->s->parent,P1=P1->s,
              P3=P2->n,P4=P3->n] exist or
             [C->e,P2=C->e->parent,
              P1=P1->e,P3=P2->w,P4=P3->w] exist ) then

      f[C] = 1/16*(-f[P1] + 9*(f[P2]+f[P3]) - f[P4]);
      (`centered' cubic polynomial interpolation)

   else if ( [C->s,P3=C->s->parent,P2=P3->s,
              P1=P2->s,P4=P3->n] exist or
             [C->e,P3=C->e->parent,P2=P3->e,
              P1=P2->e,P4=P3->w] exist ) then

      f[C] = 1/16*(f[P1] + 5*(f[P4]-f[P2]) + 15*f[P3]);
      (`backwards' cubic polynomial interpolation)

   else if ( [C->s,C->s->e,PA=C->s->e->parent,
              PB=PA->n,PC=PA->w,PD=PC->n] exist or
             [C->w,C->w->s,PA=C->w->s->parent,
              PB=PA->n,PC=PA->e,PD=PC->n] exist ) then

      f[C] = 1/4*(f[PA] + f[PB] + f[PC] + f [PD]);
      (bilinear interpolation)

   else if ( [C->s,P1=C->s->parent,P2=P1->n] exist or
             [C->e,P1=C->e->parent,P2=P1->w] exist or
             [C->w,P1=C->w->parent,P2=P1->e] exist ) then

      f[C]=1/2*(f[P1] + f[P2]); (linear interpolation)

   end if
end of subroutine interpolate
\end{verbatim}
}